\documentclass[letterpaper,titlepage,11pt]{article}

\pdfoutput=1

\usepackage{amsmath,amssymb,amsthm,mathrsfs,bbm}
\usepackage{latexsym,amscd,amsbsy,amsfonts,dsfont}
\usepackage{graphicx}
\usepackage{xcolor}
\usepackage[
      colorlinks=true,
      linkcolor=blue,
      urlcolor=blue,
      filecolor=black,
      citecolor=red,
      pdfstartview=FitV,
      pdftitle={},
        pdfauthor={Roberto Emparan, David Licht, Ryotaku Suzuki, Marija Tomasevic, Benson Way},
        pdfsubject={},
        pdfkeywords={},
        pdfpagemode=None,
        bookmarksopen=true
      ]{hyperref}\usepackage{caption}

\usepackage[utf8]{inputenc}
\usepackage{multirow}
\usepackage{makecell}

\newcommand{\beq}{\begin{equation}}
\newcommand{\eeq}{\end{equation}}
\newcommand{\beqa}{\begin{eqnarray}}
\newcommand{\eeqa}{\end{eqnarray}}
\newcommand{\bea}{\begin{eqnarray}}
\newcommand{\eea}{\end{eqnarray}}

\newcommand{\pd}{\partial}

\newcommand{\lp}{\left(}
\newcommand{\rp}{\right)}

\newcommand{\ord}[1]{{\mathcal O}\lp #1\rp}

\newcommand{\sR}{\mathsf{R}}

\def\clock{{\count0=\time
           \divide\count0 60
           \ifnum\count0<10 0\fi\the\count0
           \multiply\count0 -60 \advance\count0 \time
           :\ifnum\count0<10 0\fi \the\count0
         }}
\newcommand{\timestamp}{{\small\vbox{\hbox{\tt\jobname.tex}
\hbox{\the\day/\the\month/\the\year, \clock}}}}

\numberwithin{equation}{section}

\begin{document}

\begin{titlepage}
\rightline{\small CPHT-RR107.122021}
\rightline{\small TTI-MATHPHYS-8}
\vskip 1.3cm
\centerline{\LARGE \bf Black Tsunamis and Naked Singularities in AdS}
\bigskip

\vskip .6cm
\centerline{\bf Roberto Emparan$^{a,b}$, David Licht$^{b,c}$, Ryotaku Suzuki$^{b,d,e}$,}
\centerline{\bf Marija Tomašević$^{b,f}$, Benson Way$^{b}$
}

\vskip 0.4cm
\centerline{\sl $^{a}$Instituci\'o Catalana de Recerca i Estudis
Avan\c cats (ICREA)}
\centerline{\sl Passeig Llu\'{\i}s Companys 23, E-08010 Barcelona, Spain}
\smallskip
\centerline{\sl $^{b}$Departament de F{\'\i}sica Qu\`antica i Astrof\'{\i}sica, Institut de
Ci\`encies del Cosmos,}
\centerline{\sl  Universitat de
Barcelona, Mart\'{\i} i Franqu\`es 1, E-08028 Barcelona, Spain}
\smallskip
\centerline{\sl $^{c}$Department of Physics, Ben-Gurion University of the Negev,
Beer-Sheva 84105, Israel}
\smallskip
\centerline{\sl $^{d}$Department of Physics, Osaka City University}
\centerline{\sl Sugimoto 3-3-138, Osaka 558-8585, Japan}
\smallskip
\centerline{\sl $^{e}$Mathematical Physics Laboratory, Toyota Technological Institute}
\centerline{\sl Hisakata 2-12-1, Nagoya 468-8511, Japan}
\smallskip
\centerline{\sl $^{f}$CPHT, CNRS, Ecole Polytechnique, IP Paris, F-91128 Palaiseau, France}
\smallskip

\vskip 0.5cm
\centerline{\small\tt emparan@ub.edu, david.licht@icc.ub.edu, s.ryotaku@icc.ub.edu, }
\centerline{\small\tt marija.tomasevic@polytechnique.edu, benson@icc.ub.edu}

\vskip .9cm
\centerline{\bf Abstract} \vskip 0.2cm \noindent
We study the evolution of the Gregory-Laflamme instability for black strings in global AdS spacetime, and investigate the CFT dual of the formation of a bulk naked singularity. Using an effective theory in the large $D$ limit, we uncover a rich variety of dynamical behaviour, depending on the thickness of the string and on initial perturbations. These include: large inflows of horizon generators from the asymptotic boundary (a `black tsunami'); a pinch-off of the horizon that likely reveals a naked singularity; and competition between these two behaviours, such as a nakedly singular pinch-off that subsequently gets covered by a black tsunami. The holographic dual describes different patterns of heat flow due to the Hawking radiation of two black holes placed at the antipodes of a spherical universe. 
We also present a model that describes, in any $D$, the burst in the holographic stress-energy tensor when the signal from a bulk self-similar naked singularity reaches the boundary. The model shows that the shear components of the boundary stress diverge in finite time, while the energy density and pressures from the burst vanish.

\noindent

\end{titlepage}
\pagestyle{empty}
\small

\addtocontents{toc}{\protect\setcounter{tocdepth}{2}}

\normalsize
\newpage
\pagestyle{plain}
\setcounter{page}{1}

\section{Introduction and Summary}

Black strings in dimensions $D\geq 5$ have long been known to exhibit an instability that ripples and then clumps the horizon along its length \cite{Gregory:1993vy,Gregory:1994bj}. If the string is thin enough compared to its length, this instability causes the horizon to pinch off in finite time, forming a naked singularity \cite{Lehner:2010pn}. The process provides a broad mechanism for violating cosmic censorship in black holes with a separation of horizon length scales.

In this article, we investigate this phenomenon in anti-de Sitter space (AdS). We will see that the behavior of this system is rich and diverse, both from the viewpoint of bulk gravity and of the dual holographic conformal field theory (CFT) at the boundary.

More specifically, we study a black string in global AdS.  The fixed boundary geometry for this black string can be described as a sphere with two black holes at the antipodes. Because of their Hawking radiation, these black holes serve as heat baths for the strongly coupled boundary CFT.  Following the dynamical evolution of this radiation is a difficult problem in quantum field theory, but holography maps it to the dynamics of classical gravity in AdS.

The horizons of these boundary black holes extend into the bulk and connect to each other as a black string.  In an early study of this system, Ref.~\cite{Hirayama:2001bi} showed that when the black string horizon in the bulk is sufficiently thinner than the AdS radius (i.e., when the boundary black holes are small), its classical perturbations exhibit a rippling instability similar to that of black strings in flat Kaluza-Klein space\footnote{See also \cite{Gregory:2000gf} for the instability of black strings in Poincar\'e-AdS.}. We aim to explore the progression of this instability beyond its linearized onset, focusing on the features that lead to, or prevent, a violation of cosmic censorship. We will find qualitatively new phenomena that were not present for asymptotically flat black strings.

Of particular interest is the CFT dual to the formation of a naked singularity.  These naked singularities are pathologies in classical gravity, and are widely believed to be resolved by bulk quantum effects.  A naked singularity should therefore correspond to some pathology of the CFT in the infinite $N$ limit that would be absent at finite $N$. We attempt to understand how such a pathology manifests itself in CFT observables.

\paragraph{Thermodynamic arguments.}
In their seminal work \cite{Gregory:1993vy,Gregory:1994bj}, Gregory and Laflamme provided a simple argument for why black strings could be expected to develop a pinch that would sever the horizon.  Consider a black string: the spacetime product of a Schwarzschild black hole and a circle.  Consider also a localized black hole: a small spherical black hole in a spacetime that is asymptotically the product of flat space and a circle (i.e. Kaluza-Klein flat). Both configurations compete thermodynamically.  However, for small masses relative to the circle size, the localized black hole has higher entropy. That is, thin black strings are thermodynamically unstable in the microcanonical ensemble. This suggests (but does not prove) the existence of a dynamical instability of black strings towards localized black holes.  Such a dynamical process requires that the horizon changes topology, creating a naked singularity that violates cosmic censorship.\footnote{For an overview of the actual outcome of the instability in different dimensions, see \cite{Emparan:2018bmi}.}

We would like to make a similar thermodynamic argument for black strings in AdS. As mentioned above,  the $d$-dimensional asymptotic boundary geometry of a black string solution in global AdS$_{d+1}$ is a spatially spherical universe with two black holes at the antipodes. This geometry is described by a single dimensionless parameter, which is the radius of the black holes relative to the curvature of the sphere.

But in addition to black strings, there are other bulk solutions with the same boundary geometry. The horizons of the two boundary black holes must extend into the bulk, and can be connected either as a single bulk horizon, or disconnected as two separate horizons.\footnote{For a review of early work on these systems, see \cite{Marolf:2013ioa}. Note that we only consider the `tuned' case where the bulk horizon temperature is the same as the boundary horizon temperature. } The former are known as `black funnels', the simplest instance being the uniform black string solutions, but there may be non-uniform funnels too. Solutions with disconnected horizons are known as `black droplets'.%
\footnote{Usage of the terms `black funnel' or `black string' depends on context, even though they describe the same object in the present situation.  The funnel/droplet terminology is generally used when the AdS boundary metric contains black holes (e.g., in \cite{Marolf:2013ioa}).  The black string or localized black hole terminology refers to the analogous solutions in Kaluza-Klein flat space (e.g., in \cite{Horowitz:2011cq}).  We will use both terminologies interchangeably.} There are also configurations with more than two disconnected horizons, e.g., droplets with a spherical black hole between them. The uniform funnel has an exact solution, but other configurations are either known only numerically or merely expected to exist by intuitive or semi-analytic arguments.  In general, many of these compete with each other thermodynamically.

To determine which solutions are dominant, we need to first choose a thermodynamic ensemble.  Unlike the situation in Kaluza-Klein flat space, the bulk black holes are non-compact.  On the CFT, the boundary black holes function as heat baths that are allowed to exchange heat and energy with the CFT.  We should therefore work in the canonical ensemble where only solutions with the same boundary metric and the same bulk temperature compete with each other. In this ensemble, configurations with the lowest free energy are preferred.  Dynamically, the black hole area theorem is replaced by an analogous statement that the free energy monotonically decreases \cite{Bunting:2015sfa}.

Ref.~\cite{Marolf:2019wkz} carried out an extensive numerical study of several of these solutions.  Their results indicate that uniform funnels can dominate the thermodynamics when they are very thick (i.e., when the boundary black holes are large). This conclusion is further supported by the perturbative stability of thick AdS black strings \cite{Hirayama:2001bi}, and is analogous to the dominance of thick black strings on a Kaluza-Klein circle. However, unlike the Kaluza-Klein case, the droplet phases, which are analogous to the localized black holes, are never thermodynamically preferred. Instead, when the boundary black holes are small, the preferred phases are highly non-uniform fat funnels, which have much lower free-energy than droplets.  These fat funnels roughly resemble a large bulk black hole with two small connections to the AdS boundary.  The fact that fat funnels are preferred agrees with the fact that at high temperatures, large black holes in AdS are preferred over small ones or over empty (thermal) AdS.

At face value, this thermodynamic analysis suggests that thin, unstable AdS black strings should evolve directly into fat funnels, which would avoid any violation of cosmic censorship.

\paragraph{Black tsunami and naked pinch.}

The evolution of a thin string towards a fat funnel requires that the string fattens with a dramatic increase in its entropy. This is indeed possible, since the bulk horizon is non-compactly generated, and can grow indefinitely by the inflow of null generators from the asymptotic boundary. We refer to this large inpouring phenomenon as a `black tsunami'.

In the dual field theory, the fixed, non-dynamical boundary black holes act as infinite sources (or sinks) of heat for the CFT, and can therefore flood the surrounding universe with large amounts of radiation. But despite the presence of these black holes, the initial CFT state dual to the AdS black string is, quite peculiarly, unexcited without any thermal radiation component \cite{Gregory:2008br}. It seems natural that when the size of the black holes, and the wavelength of their radiation, are much smaller than the radius of the universe, the black holes will pour out CFT radiation until thermal equilibrium is reached.\footnote{It may seem that this setup is a dynamical means of realizing the Hawking-Page transition \cite{Hawking:1982dh,Witten:1998qj}, with the boundary black holes serving both as heat baths and as nucleation sites that can trigger a deconfinement phase transition dynamically. However, the initial state has a horizon that stretches all along the bulk and presumably should not be regarded as a confining phase. A droplet phase, though, could nucleate the deconfinement transition.}

This dynamical growth of the bulk horizon can plausibly account for the eventual dominance of fat funnels over smaller droplet phases as suggested by bulk thermodynamics. However, the conclusion that the evolution will always avoid a singular pinch cannot be right. Consider a uniform black string that is much thinner than the AdS radius.  Near the center of AdS, the black string resembles a Ricci-flat black string, so we expect that its horizon will generically pinch off in a time scale set by its thickness. This is much too quick for the thin string to be affected by the AdS geometry, which occurs at a time set by the AdS length. The singularity therefore forms before the string has had time to grow fat through inflow from the boundary.

\paragraph{Evolution scenarios.}

In order to study the dynamical evolution of unstable AdS black strings, we use an effective theory in a $1/D$ expansion of the type developed in \cite{Emparan:2015gva} (see also \cite{Emparan:2016sjk,Dandekar:2016jrp}, and \cite{Emparan:2020inr} for more references). Our results indicate a rich pattern of possible dynamical behaviour, depending mostly on the thickness of the initial black string and on the perturbations applied to it. We place each evolution into one of three broad categories:

\begin{enumerate}
\item Black tsunami, no pinching; direct evolution to fat funnel.

\item Direct pinch-off to droplet-like configuration, or direct evolution towards a non-uniform string.

\item Competition between the first two scenarios.  For example, evolution towards a pinch-off, later dominated by a tsunami, ending at a fat funnel.
\end{enumerate}

\begin{figure}[h]
    \centering
    \includegraphics[width=0.9\textwidth]{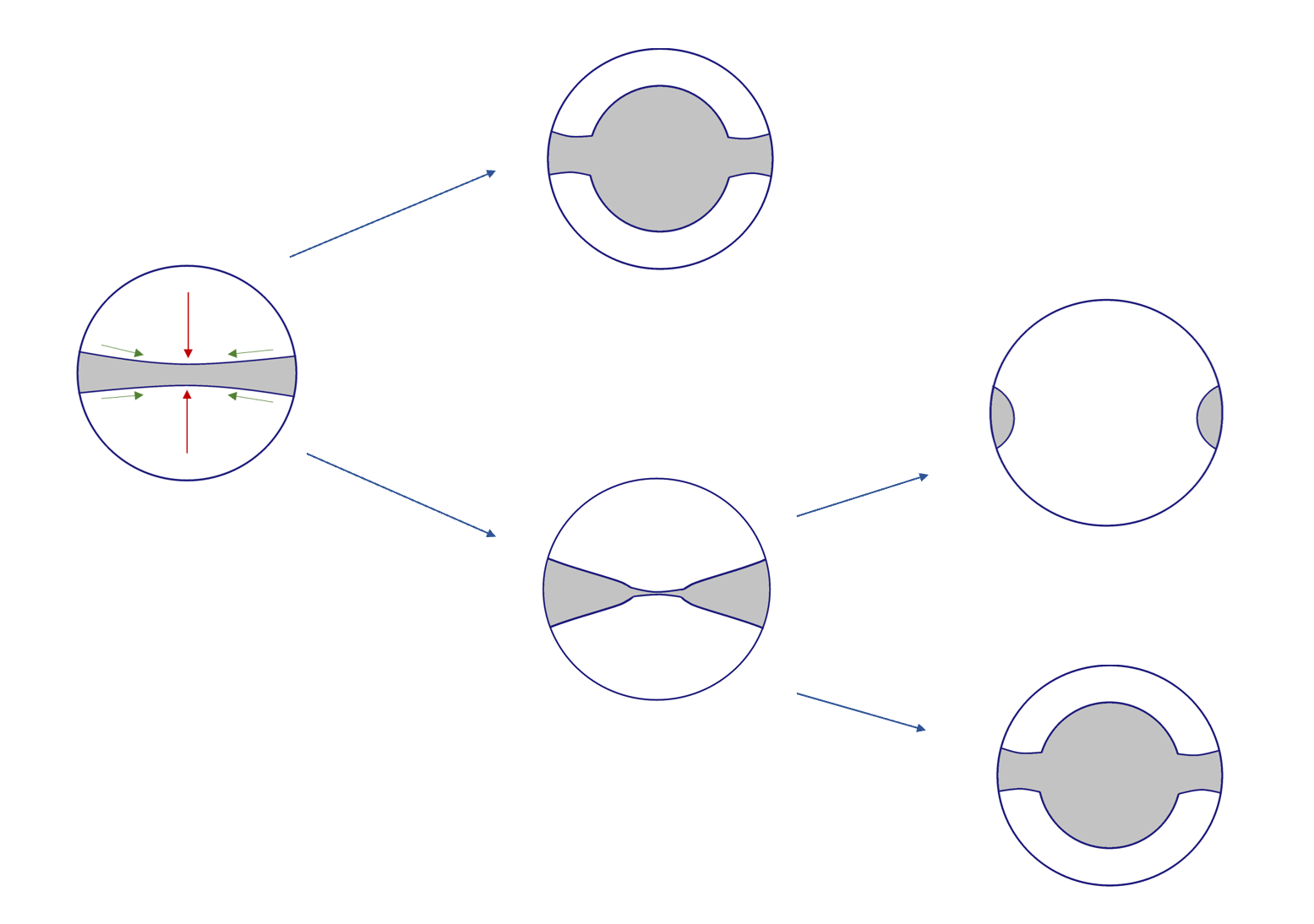}
    \caption{\small Representatives of three scenarios for the evolution of unstable black strings in AdS. The two competing effects, shown in the left diagram, are the pinching near the center (red arrows) and the inflow of generators from the boundary, dubbed black tsunami (green arrows). In the first scenario (top), the tsunami dominates and the system is directly driven to a fat black funnel. In other scenarios, the horizon pinches, leading to a naked singularity that splits the horizon into twin black droplets without triggering a tsunami (middle); or the pinch forms, but is later covered by a black tsunami, ending again at a fat funnel (bottom). Other possible final configurations that are not shown include metastable non-uniform strings, or central black holes between droplets.}
    \label{fig:evolution}
\end{figure}

Let us elaborate more on these alternatives.  First, when the uniform string is unstable, the dominant thermodynamic phase corresponds to a fat funnel. Scenario 1 is simply the evolution directly towards this preferred phase. Our results indicate that the size of the central bulge is larger than the funnel neck by a factor that diverges with $D$.

However, there are a number of non-uniform strings that are dynamically stable, but are not the thermodynamically preferred state, and are in this sense metastable.  The thinner the string, the more of these non-uniform solutions exist as possible endpoints to evolution. Scenario 2 is the direct evolution towards one of these states, with the non-uniformity possibly becoming very large.  Strictly speaking, pinch-off never occurs in the infinite $D$ effective theory, but the horizon can become arbitrarily thin.  At finite $D$, the evidence in \cite{Emparan:2018bmi} indicates that some of these highly non-uniform strings will correspond to a pinch-off. This case evolves towards a droplet configuration with the appearance of a naked singularity.

The dynamics in scenario 2 can involve the competition of several unstable modes, and can be fairly complex.  There are several variations within this scenario that we have found.  One is a simple direct evolution towards a non-uniform solution, possibly one resembling droplets.  Another resembles the formation of a central black hole between two droplets.  In this latter case, at infinite $D$ the central black hole is still connected to the droplets by a thin horizon; the configuration is long-lived, but eventually settles towards a non-uniform string. If a pinch-off occurs at finite $D$, it can leave the central black hole between the droplets in a stable position at the center of AdS.

Finally, scenario 3 involves the competition between instabilities that drive the evolution towards a fat funnel, and others that drive it towards pinches.  For example, we have observed situations where the horizon quickly draws back from the center, while a slower, massive horizon wave comes from the boundary, and eventually engulfs the central region to form a fat funnel. This process resembles the manner in which the sea water recedes from the shoreline and exposes the seafloor ahead of the arrival of a tsunami.  At finite $D$, this should correspond to a pinch-off of the horizon at early times, which later gets washed out in the formation of a fat funnel.

As we see in several scenarios, the evolution can involve competing modes and is therefore difficult to predict in advance.  In general, thinner strings have more unstable modes and more possible end states, and therefore their evolution can be more complex.

\paragraph{Dual CFT: floods, halos, and shearing bursts.} Translating these qualitative bulk evolutions into dual boundary terms is straightforward, but the conclusions are surprising. The black tsunami is dual to the expected flooding of the compact boundary universe with radiation emitted by the hot black holes.  On the other hand, the bulk horizon pinch-off causes a non-thermal burst of radiation in the CFT which does not issue from the (boundary) black holes, and which lasts only for a time set by the black hole radius. In some cases, this burst will be followed by a slower tsunami of radiation plasma that fills up the space. In other cases (dual to stable droplet formation) there will be no flood of radiation, which will instead remain in a halo of plasma around the black holes.

On general grounds, the burst will consist of a very high frequency signal in the stress-energy tensor. Note, though, that if the singularity were continuously self-similar, then a simple argument, given below, shows that it would leave no strong signal in the one-point functions at the boundary (although other, less local probes may be sensitive to it).

To further find out the structure of the burst, we will generalize a model presented in \cite{Chesler:2019ozd} for the formation of a discrete self-similar naked singularity in the AdS bulk. The model suggests that this causes a signal at the boundary that diverges at finite time, and this was verified in a non-linear numerical analysis of critical collapse \cite{Chesler:2019ozd}. We find that in the gravitational field extension of the model not all the components of the boundary stress-energy tensor behave in the same manner. The shear stresses diverge  as $1/(t-t_*)$, while the momentum remains constant and the energy density and pressures vanish as $(t-t_*)$. 

Even without a divergence in some of the stress-energy components, the boundary signal is not smooth, as the oscillations become infinitely rapid due to discrete self-similarity. Indeed, this is the reason that the shear grows large even if the energy density vanishes. 

Ultimately, the evolution onto a Planckian curvature region in spacetime should naturally excite quanta with Planck-scale energies. In the dual CFT, these are excitations with energy densities that scale with $N^2$. Our result that the amplitude of the energy density signal is vanishingly small indicates that there will be only a few---less than $\ord{N^2}$, and presumably $\ord{1}$---of these quanta, which, given their rapid oscillation rate, will create large but localized shears.

\medskip

We now proceed as follows. In the next section, we introduce the large-$D$ effective formalism that allows us to non-linearly evolve the system. Sec.~\ref{sec:evolutions} describes the results of our numerical solutions. In Sec.~\ref{sec:signal}, we change gear and present a model (in finite $D$) for the propagation to the AdS boundary of the signal of a self-similar singularity in the bulk. Sec.~\ref{sec:outlook} contains further discussion of our results. Appendix~\ref{app:gravss} provides the details of the calculations for Sec.~\ref{sec:signal}.

\section{Setup}
\label{sec:setup}

Our initial configuration is the solution for a black string in AdS in $D=n+4$ dimensions, which in Eddington-Finkelstein coordinates takes the form
\beq\label{adsbs}
ds^2= \frac{L^2}{\cos^2 z}\left(dz^2 -f(r)dv^2+2 dv dr+r^2 d\Omega_{n+1}\right)\,,
\eeq
with
\beq
f(r)=r^2+1 -\left(\frac{r_0}{r}\right)^n\,.
\eeq
Every slice of constant $z\in(-\pi/2,\pi/2)$ is a Schwarzschild-AdS$_{n+3}$ geometry, so this describes a black string with a horizon that extends along the $z$ direction. At the boundary of the spacetime, at $z\to\pm\pi/2$, two identical Schwarzschild-AdS$_{n+3}$ geometries  are glued together at their $(n+2)$-dimensional conformal boundaries to form a static, compact universe. In an appropriate conformal frame, the spatial geometry of this universe is essentially a sphere with two black holes at the antipodes.

The horizon coordinate radius $r_H$ is the largest positive root of
\begin{equation}
    r_H\left(1+r_H^2\right)^{1/n}=r_0\,,
\end{equation}
and its physical size is measured in units of the bulk cosmological radius $L$, which scales out of all our equations (so we could set $L=1$). We will employ $r_0$ as the parameter of the system.

It was shown in \cite{Hirayama:2001bi} that thin $D=5$ black strings with $r_0\lesssim 0.20$ are unstable to small, linearized perturbations. As we will see, in the limit when $D$ is large, the instability is present when $r_0<1/\sqrt{2}$ \cite{Marolf:2019wkz}. The evolution of this instability beyond the linear perturbation regime has not been studied in any $D$ yet, though work is in progress \cite{GLAdSnumerics}. Here we will follow its progression using an effective theory of black hole horizons in the limit of large $D$ similar to the one in \cite{Emparan:2015gva}. These theories have proven efficient for studying the dynamics of complex black hole systems that involve horizon pinches \cite{Emparan:2018bmi,Andrade:2018yqu,Andrade:2019edf,Andrade:2020ilm}.

\subsection*{Effective theory at large $D$}

We take the large $D$ (i.e., large $n$) limit of these spacetimes, focusing on the region near the center of AdS, since it is here that most of the dynamics of the strings occurs. Let us illustrate this limit for the uniform black string  \eqref{adsbs}. We introduce new coordinates
\begin{equation}
    x=\sqrt{n} z\,,\qquad \sR =\left(\frac{r}{r_0}\right)^n
\end{equation}
which, when kept finite as $n\to\infty$, localize the geometry near $r=r_0$ (and hence near $r=r_H=r_0+\ord{1/n}$) and near the AdS center at $z=0$. Further rescaling, for convenience, $v=t/r_0$, the black string solution \eqref{adsbs} becomes
\beq\label{bslargen}
ds^2 =\frac{L^2}{\cos^2(x/\sqrt{n})}\left( \frac{dx^2}{n}-\left( 1+r_0^{-2}\right)\left( 1-\frac{m_0}{\sR}\right) dt^2+\frac2{n}\frac{dt d\sR}{\sR} +r_0^2\sR^{2/n} d\Omega_{n+1}\right)
\eeq
with constant $m_0=(1+r_0^2)^{-1}$. This motivates the following ansatz for the metric,
\beq
ds^2=\frac{L^2}{\cos^2(x/\sqrt{n})}\left( \frac{H}{n}dx^2-\left( 1+r_0^{-2}\right)A dt^2+u_t \frac2{n} \frac{dt d\sR}{\sR}-\frac{2}{n}Cdt dx +r_0^2\sR^{2/n} d\Omega_{n+1}\right)\,,
\eeq
where $H$, $A$, $C$ and $u_t$ are functions of $(t,x,\sR).$
Then, the $\sR$-dependence in the Einstein-AdS equations, expanded to leading order in $1/n$, can be explicitly integrated giving
\begin{align}\label{eq:ACHu}
A&=1-\frac{m(t,x)}{\sR}\,,& C&=\frac{p(t,x)}{\sR}\,,\\
H&=1+\frac{p^2(t,x)}{n\lp 1+r_0^{-2}\rp m(t,x)\sR}\,,& u_t&=1-\frac{p^2(t,x)}{2n\lp 1+r_0^{-2}\rp m(t,x)\sR}\,,
\end{align}
and the full set of Einstein-AdS equations are solved when the functions $m(t,x)$ and $p(t,x)$ satisfy
\beq\label{meqn}
 \pd_t m + \lp \pd_x +x \rp \lp p-\pd_x m\rp=0\,,
\eeq
\beq\label{peqn}
\partial_t p-\lp \pd_x +x \rp \lp\pd_x p-\frac{p^2}{m}\rp + p-\left(1+\frac1{r_0^{2}}\right)\pd_x m=0\,.
\eeq
These two equations provide us with an effective theory of the dynamics of horizons in AdS.

We note that the equations are invariant under the scaling symmetry $m\to\lambda m$, $p\to\lambda p$.  When we study dynamical evolution, we will fix this scaling symmetry with an appropriate boundary condition.

\subsection*{Exact solutions and linear instability}

A detailed study of the derivation of these equations, their solutions and their properties, will be given elsewhere \cite{GLAdSlong}. Here we will only discuss the main features, starting from their simplest solution, namely, the black string
or uniform funnel \eqref{bslargen},
\beq
m=m_0\,,\qquad p=0\,.
\eeq

To study its linear stability, we look for normalizable perturbations of the form\footnote{The gaussian factors ultimately originate from $\cos^n(x/\sqrt{n})\simeq e^{-x^2/2}$.}
\begin{equation}\label{linpert}
m(t,x)= m_0+\epsilon\, e^{\Omega t} f(x) e^{-x^2/2}\,,\qquad p(t,x)=\epsilon\, e^{\Omega t} g(x) e^{-x^2/2}\,,
\end{equation}
where $f(x)$ and $g(x)$ are finite polynomials. When we plug \eqref{linpert} in the effective equations \eqref{meqn} and \eqref{peqn}, the requirement that the polynomials truncate at finite order determines that the lowest mode, with
\begin{equation}
    f=1\,,\qquad g=-(1+\Omega)x\,,
\end{equation}
has a growth rate
\begin{equation}
    \Omega=-2\pm \sqrt{2+r_0^{-2}}\,.
\end{equation}
This mode grows exponentially when
\begin{equation}\label{instabbound}
    r_0<\frac1{\sqrt{2}}\,.
\end{equation}
This is then the instability regime for AdS black strings in the limit $n\to\infty$. More generally, there exist linearized modes that become unstable when $r_0$ is smaller than a stability threshold at
\begin{equation}\label{zeromoder0}
    r_0=\frac1{\sqrt{2+k}} \,,\qquad k=0,1,2,\dots
\end{equation}
For these values of $r_0$ we find zero modes with
\begin{equation}
    f(x)=H_k(x/\sqrt{2})\,,\qquad g(x)=-\frac1{\sqrt{2}}H_{k+1}(x/\sqrt{2})
\end{equation}
(where $H_k$ are Hermite polynomials) which solve the linearized equations with $\Omega=0$.

Another exact solution of \eqref{meqn} and \eqref{peqn} is the static `gaussian blob'
\beq\label{gblob}
m=m_0\, 
e^{-\lp 1+r_0^{-2}\rp x^2/2 }\,,\qquad p=\partial_x m\,.
\eeq
In this case, the mass density $m$ approaches zero at $x\to\pm\infty$. The interpretation of such localized blobs is the same as in the asymptotically flat `blobology' of \cite{Andrade:2018nsz}: these are the effective-theory description of large-$D$ Schwarzschild-AdS black holes. Indeed, it is easy to show that these blobs can be recovered by appropriately taking the limit $D\to\infty$ of the known, exact solution for Schwarzschild-AdS$_D$ black holes. As expected from this connection, these blobs can be proven to be dynamically stable.

A few other solutions of the effective theory can be found analytically, but many others require a numerical solution of the equations, including non-uniform solutions that branch from the zero modes. Configurations that have a `ditch' along which $m$ becomes exponentially small can represent black droplets.

\subsection*{Thermodynamic stability}
Since the boundary conditions for the system do not conserve energy (which can flow in or out from $x=\pm\infty$), the quantity that determines which phase is thermodynamically preferred is not the entropy but the free energy $\mathcal{F}$. We will be considering configurations that extend to infinity with non-zero $m$, for which we define the value of $\mathcal{F}$ relative to the uniform string with $m(x)=m_0$. It is shown in \cite{GLAdSlong} that this free energy is given by
\begin{equation}
 \Delta {\cal F}(t) =\frac{r_0^2}{r_0^2+1} \int_{-\infty}^\infty dx\, e^{\frac{x^2}{2}}\left(\frac{(\partial_x m)^2}{2m}+\frac{1}{2}mv^2-(r_0^{-2}-1)(m\log (m/m_0)-m+m_0)\right)\,,
\end{equation}
where we have introduced the velocity $v(t,x)=(p-\partial_x m)/m$.

The field equation \eqref{peqn} implies that
\begin{equation}
	\frac{d\Delta {\cal F}}{dt} = -\frac{2r_0^2}{r_0^2+1}\int_{-\infty}^\infty  dx\,e^\frac{x^2}{2}m (\partial_x v)^2 \leq 0,
\end{equation}
which shows that $\Delta {\cal F}$ monotonically decreases due to viscous dissipation of the expansion $\partial_x v$. This is the statement of the second law of thermodynamics in this system.

The monotonicity of this functional is a useful guide for understanding the fate of the instability of the string. For the gaussian blob \eqref{gblob} its total value
\begin{align}
    \Delta {\cal F} = ({\rm finite \ terms}) + \frac{r_0^2-1}{r_0^2+1}m_0 \int_{-\infty}^\infty  dx\,e^\frac{x^2}{2}
\end{align}
diverges since the boundary values are different than for the string. But we can still extract relevant information from this formal expression, since the overall sign of $\Delta {\cal F}$ changes with $r_0$.
For thin strings with $r_0<1$, the gaussian blob has negative infinite free energy compared to any string, uniform or non-uniform. Hence, growing a large blob is favored for thin strings. For fatter strings, with $r_0>1$, the uniform string is instead thermodynamically favored. The fact that this bound on thermodynamic instability differs from the linear instability dynamical bound \eqref{instabbound} means that uniform strings in the range $1/\sqrt{2}<r_0<1$ are metastable, i.e., are stable for infinitesimal or small perturbations, but unstable for sufficiently large perturbations. For $r_0>1$, we expect them to be absolutely stable.

\subsection*{Fat funnels}

In the following we study configurations such that $m(t,\pm\infty)=1$, which fixes the scaling symmetry of the equations and guarantees that we have a black hole of finite radius at the AdS boundary, with a size that, in units of $L$, is controlled by the parameter $r_0$.

With these boundary conditions, there are several distinct families of funnel solutions at large $D$.  We call these ``fat funnels" and ``moderately overweight funnels."  The latter are metastable non-uniform black string solutions with a central bulge.  The former, however, are stable funnels that have sizes much larger than $L$.

Fat funnels have a size in units of $L$ that grow with $D$ and would seem to lie outside the proper remit of the large-$D$ effective theory. However, if we rescale the size of the solution such that the central bulge remains finite as $D\to\infty$, the neck size will shrink to zero.\footnote{We may consider that the AdS radius $L$ is rescaled by the same factor of $D$ as the size of the funnel bulge.} In this case, we expect that the gaussian blob solution \eqref{gblob}, which has  $m(t,\pm\infty)=0$, is a good description of these rescaled fat funnels.  This interpretation indeed seems appropriate in the light of some of the evolutions that we will present below.

\section{Dynamical evolution}
\label{sec:evolutions}

We have followed the dynamical evolution of unstable black strings with different values of $r_0<1/\sqrt{2}$, numerically solving \eqref{meqn} and \eqref{peqn} with initial data corresponding to perturbations that excite different modes of the uniform solution. Specifically, we excite a $k$-mode by setting
\begin{equation}
    m(0,x)=\exp\left(\epsilon\, e^{-x^2/2}H_{k}(x/\sqrt{2}\right)\,,\qquad p(0,x)=-m(0,x)\frac{\epsilon}{\sqrt{2}} e^{-x^2/2}H_{k+1}(x/\sqrt{2})\,,
\end{equation}
with $\epsilon$ a small parameter. When $r_0$ takes the value \eqref{zeromoder0} and $\epsilon$ is infinitesimal, the configuration is static.  For other values of $r_0$, the instabilities kick in and time evolution promptly follows.

We work with a half-domain that imposes the appropriate reflection symmetries at $x=0$, and the asymptotic boundary conditions are $m(t,\infty)=1$, $p(t,\infty)=0$.

For numerical implementations, we find it convenient to set $m=e^\mu$, $p=\Pi e^\mu$, and work with $\mu(t,x)$ and $\Pi(t,x)$ instead.  For spatial discretization, we use pseudospectral methods with a Chebyshev half-grid on the coordinate $X=\tanh(x/\sqrt 2)$.  We take time steps using the classic Runge-Kutta method RK4.

\begin{figure}[th]
    \centering
    \includegraphics[width=0.45\textwidth]{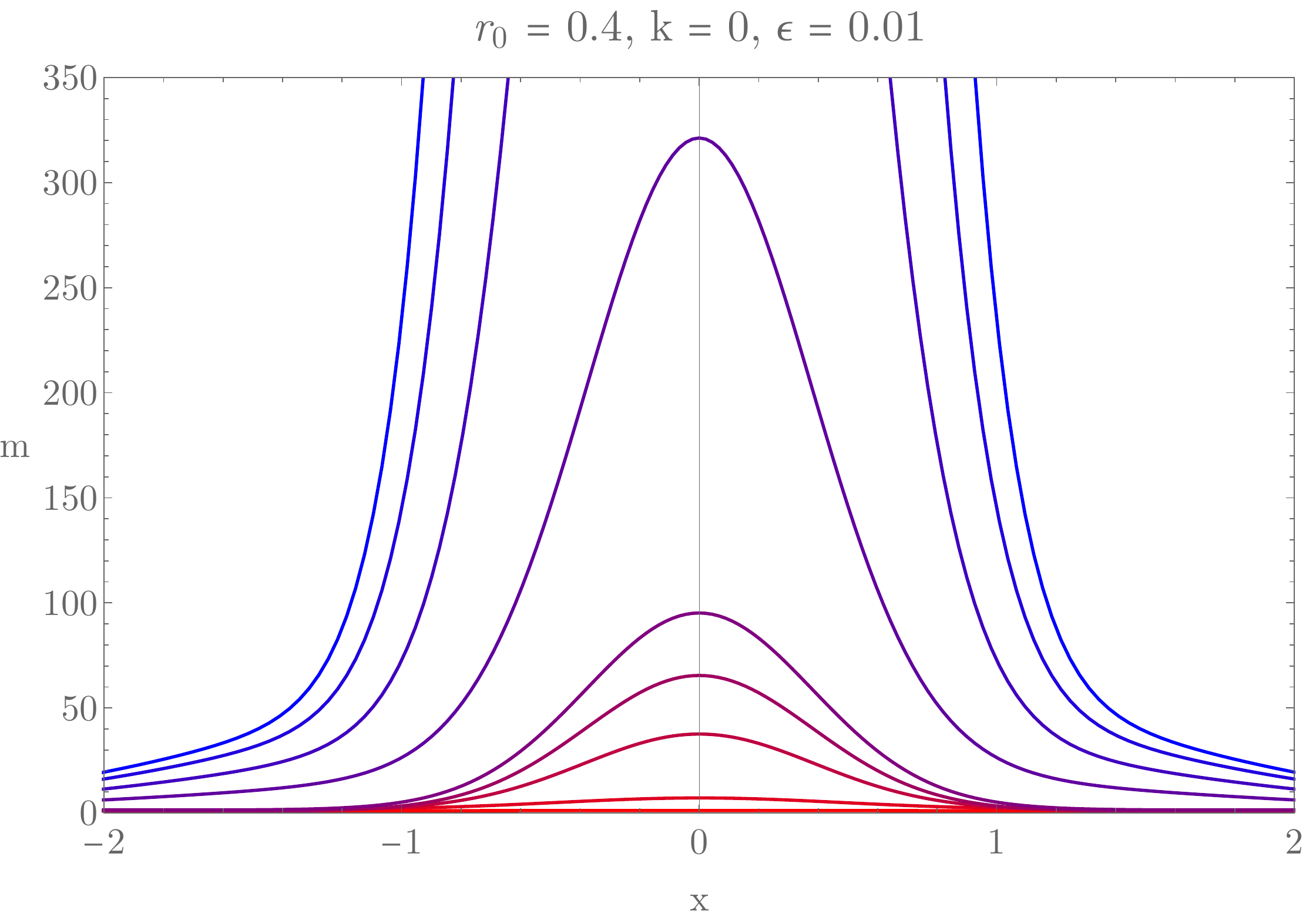}
    \includegraphics[width=0.45\textwidth]{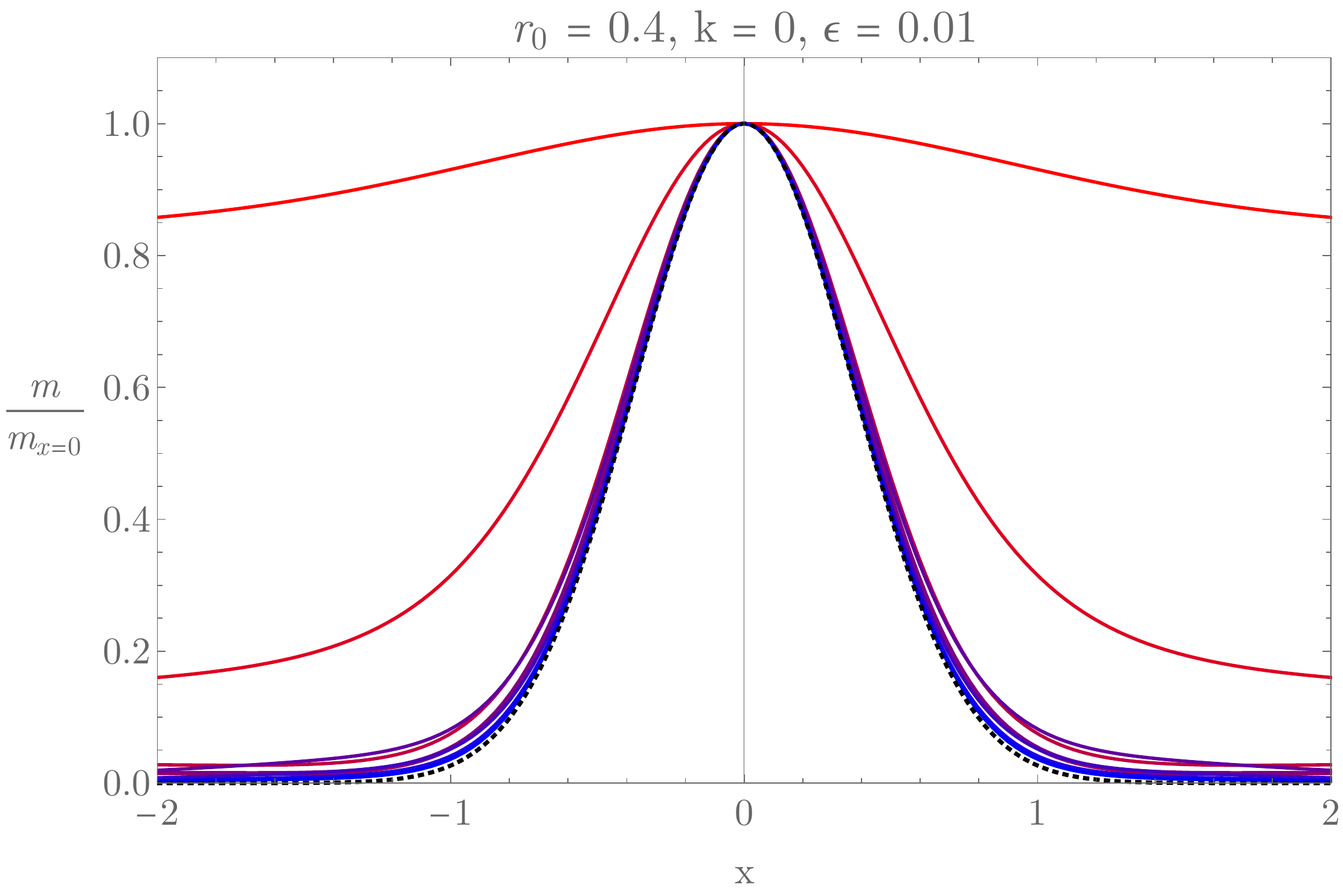}
    \caption{\small Scenario 1. Left: Indefinite growth of a tsunami, with each higher curve a time $t\sim 3.33$ later than the previous one. Right: Same as left, but with mass function rescaled by its value at $x=0$, showing approach to the gaussian blob (dashed line). This is interpreted as evolution to a very fat funnel.}
    \label{fig:growth}
\end{figure}

With some exploration of different initial conditions (changing the initial thickness $r_0$, the perturbation mode number $k$, and the sign of the small amplitude parameter $\epsilon$), it is not difficult to identify the three main scenarios described in the introduction. We present illustrative examples of each. Fig.~\ref{fig:growth} corresponds to scenario 1: the black string develops a central bulge that grows indefinitely without bound.  At late times, after rescaling the amplitude to have $m(0)=1$, we see that the bulge is well approximated by the Gaussian blob \eqref{gblob} with unit central amplitude $m_0=1$. As discussed above, this means that the infinitely-growing solution looks more and more like a large black hole, approximating a fat funnel in which the central bulge is much larger than the neck, in fact infinitely larger in the limit $D\to\infty$. Therefore, we conclude that the evolution drives the system to a very fat funnel.

\begin{figure}[th]
    \centering
    \includegraphics[width=0.45\textwidth]{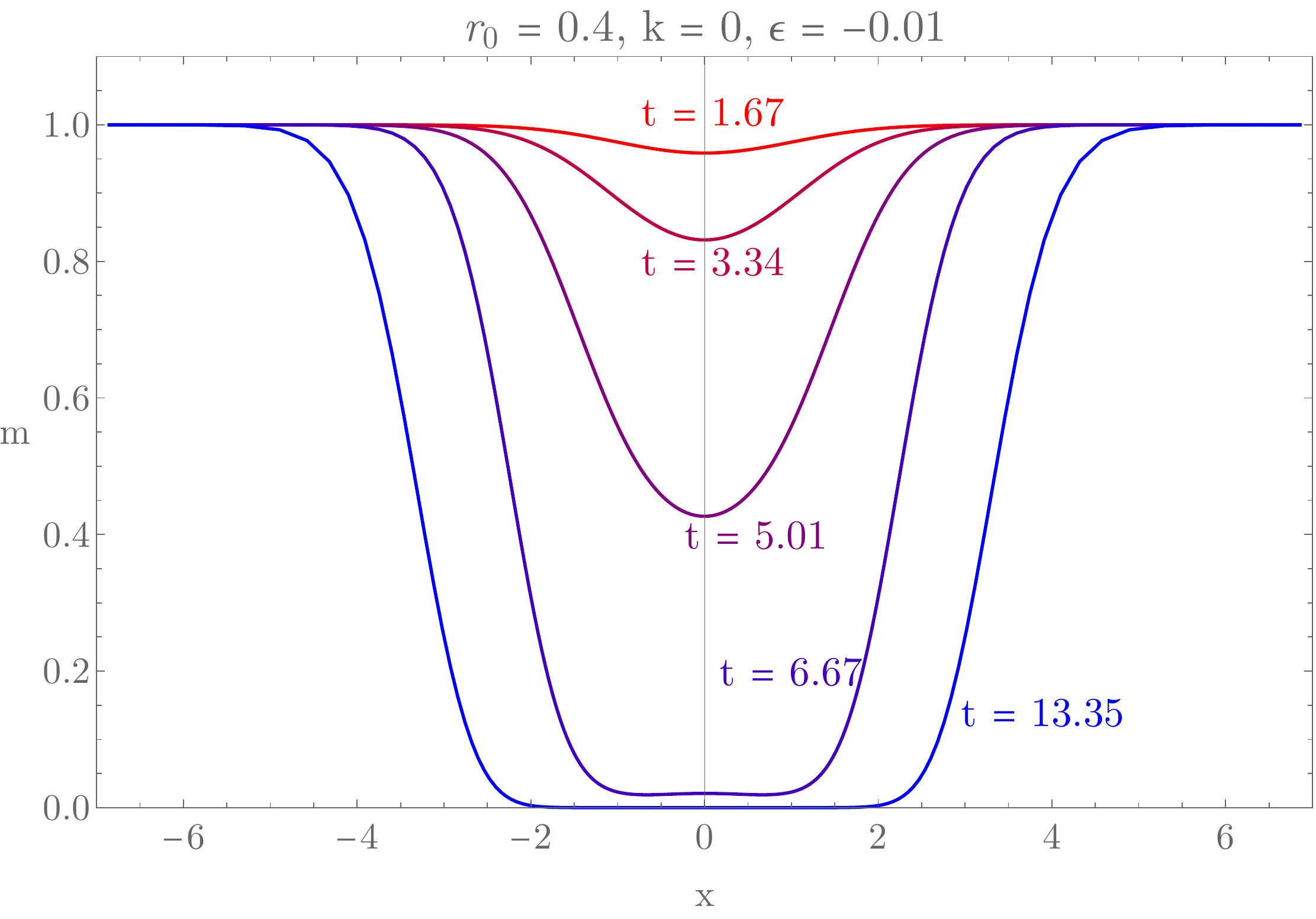}
    \includegraphics[width=0.45\textwidth]{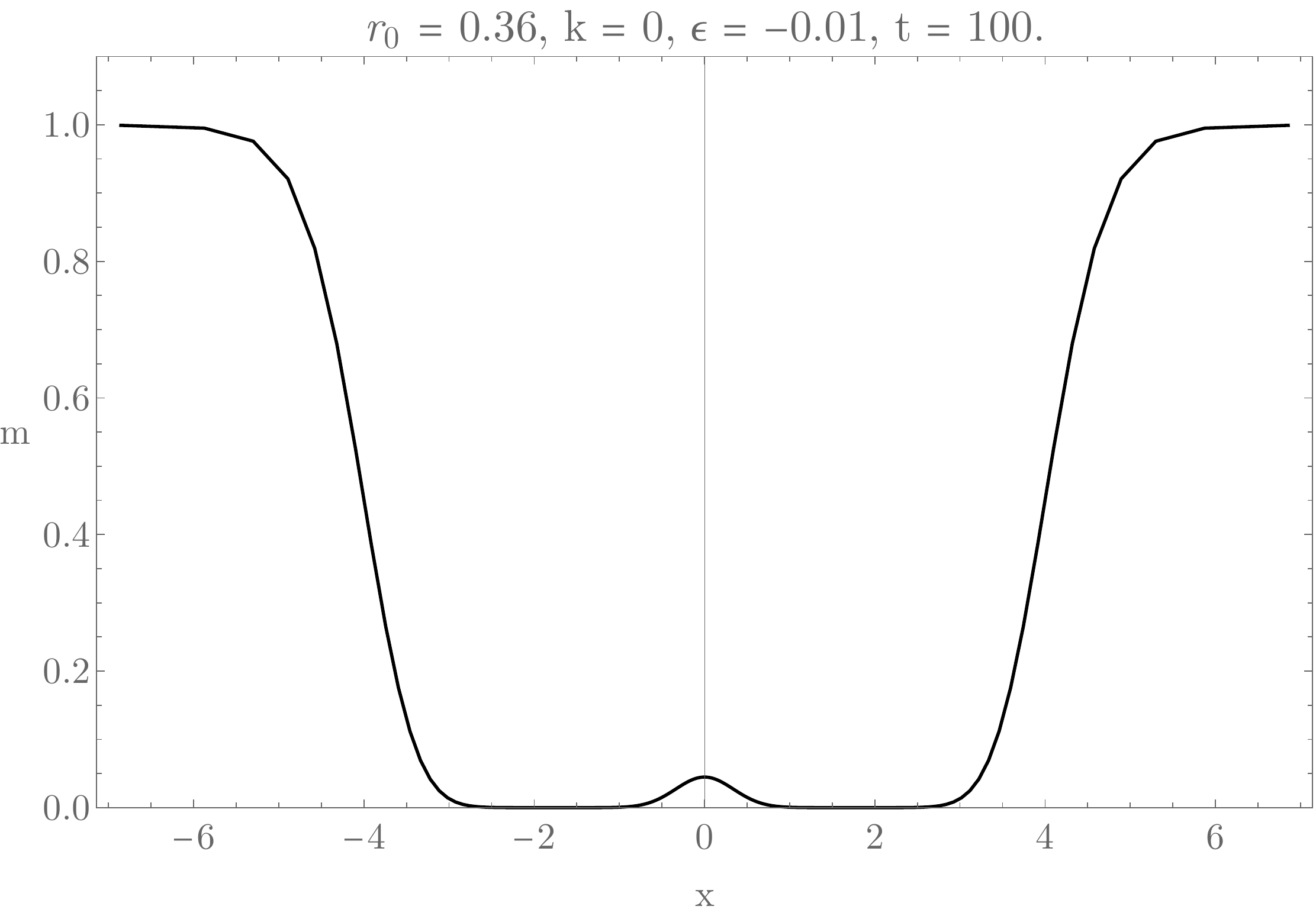}
    \caption{\small Scenario 2. Left: Pinch to a final static configuration with a finite ditch, corresponding to a twin black droplet. The difference with the evolution in Fig.~\ref{fig:growth} is the sign of the initial perturbation, which here pushes downwards. Right: Formation of a small central transient black hole, which very slowly decreases in size. The initial string is thinner than in the previous case, and therefore contains more unstable modes.}
    \label{fig:ditch}
\end{figure}

Outcomes in scenario 2 are illustrated in Fig.~\ref{fig:ditch}. In the left we see that a dip forms at the center of the string, which eventually stabilizes in a ditch of finite extent in the $x$ direction. Notice that, contrary to what happens in the tsunami, this requires a flow towards $x\to\pm\infty$ and hence the absorption of horizon generators at the boundary.

These final ditch configurations are well approximated by static solutions of the effective theory that are stable to small perturbations. At infinite $D$, these solutions are just (highly) non-uniform strings.  However, depending on $r_0$, the thickness of the string, i.e., the horizon size, near $x=0$ can become arbitrarily small.  If the horizon is sufficiently small, then it is natural to expect that at finite $D$ it will pinch off, leading to a naked singularity that is mild, in the sense of having small extent and small mass \cite{Emparan:2020vyf}. If the singularity is resolved, e.g., by quantum gravity, then the horizon will split and change its topology. Perhaps the most striking example where the large-$D$ effective theory correctly predicted pinch-off and cosmic censorship violation at finite $D$ (down to $D=6$) is the collision of higher-dimensional black holes with enough total angular momentum \cite{Andrade:2018yqu,Andrade:2019edf,Andrade:2020dgc}. This gives us confidence that, in the present case, the ultimate configuration will resemble two droplets.

It is also easy to find situations where a small blob forms in the middle of the ditch (Fig.~\ref{fig:ditch}, right), corresponding to a small black hole. In this case, the central black hole slowly decreases in size, so the final state is likely a ditch.  In the large $D$ effective theory, the central black hole is slowly leaking its energy to the outer two droplets through a very small, string-like horizon.  At finite $D$, one would expect a configuration corresponding to two droplets with a central spherical black hole between them.  Due to the severed horizon connection, it is possible that this configuration is static and lasts indefinitely as a metastable final state.

\begin{figure}[th]
    \centering
    \includegraphics[width=0.45\textwidth]{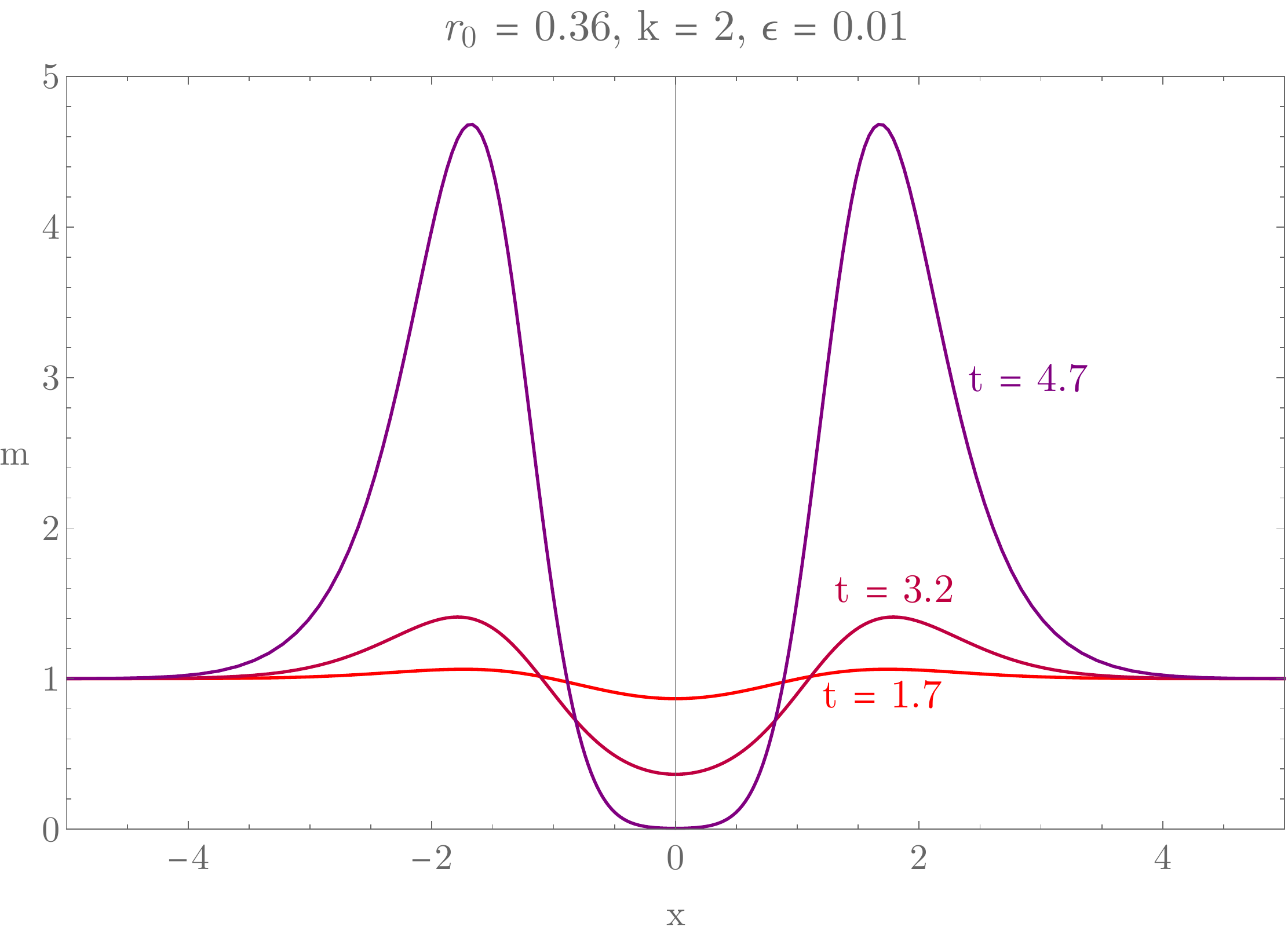}
    \includegraphics[width=0.45\textwidth]{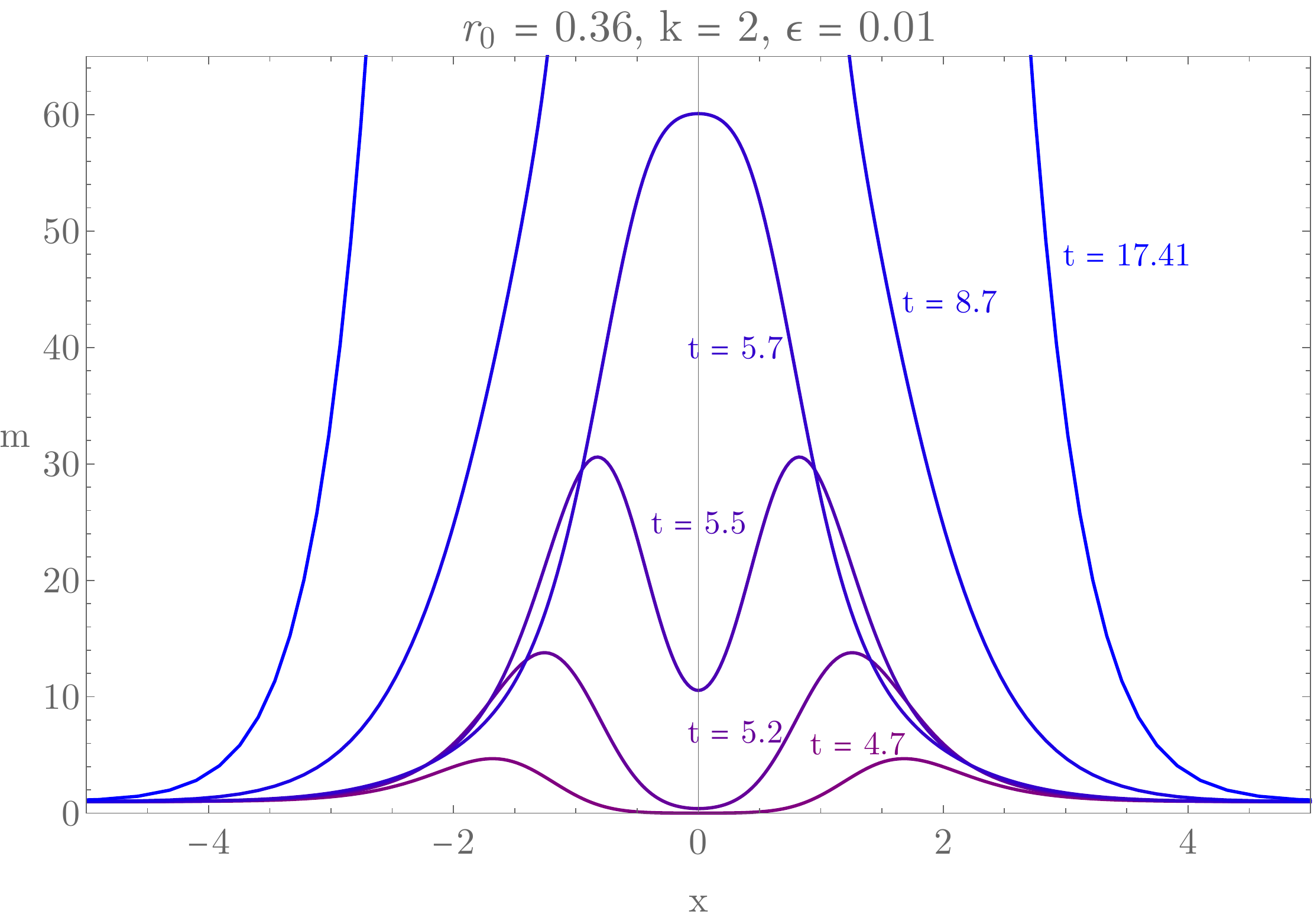}
    \caption{\small Scenario 3. Left: initial formation of a pinch. Right: subsequent evolution, zoomed out in the vertical axis to show the unbounded growth after the arrival of the black tsunami. The initial perturbation is a higher mode than in the previous figures, and manages to trigger both the pinch and the tsunami.}
    \label{fig:tsunami}
\end{figure}

Finally, in Fig.~\ref{fig:tsunami} we show scenario 3. We see (left) that the initial growth of the black string is accompanied by the development of a pinch at the center of the black string. Once again, in the effective theory at $D\to\infty$ this pinch can never reach zero size, but when $D$ is finite but sufficiently large it is expected to result in a naked singularity. Since the singularity is mild, it should not hinder the further evolution of the system, which, as Fig~\ref{fig:tsunami} (right) shows, continues to grow due to the tsunami from the boundary. Eventually, the central region is again covered by a horizon (with $m(t,0)\neq 0$) and proceeds towards an end state of the same kind as in scenario 1.

Given initial data, it is generally difficult to predict which scenario describes the evolution without explicitly doing the calculation.  But more can be said for specific parameter ranges.  As we have mentioned, the uniform black string is absolutely stable and is likely the dominant phase for $r_0>1$.  In the range $1/\sqrt 2<r_0<1$, the uniform string is metastable.  Small perturbations will return to the uniform configuration, but sufficiently large ones will provoke a black tsunami, as in scenario 1.  In the range $1/2<r_0<1/\sqrt 2$, there is a single unstable mode that grows exponentially to dominate the dynamics.  Depending on the sign of this mode perturbation, evolution either proceeds directly to a black tsunami (as in scenario 1) or to a metastable ditch (as in scenario 2).  For $r_0<1/2$, there are multiple competing unstable modes, and the dynamics becomes complicated.

\section{Boundary CFT signal of naked singularities}
\label{sec:signal}
We have given evidence that the dynamics of black strings in AdS can lead to violations of cosmic censorship.  What is the CFT dual to these naked singularities?  The pathological behaviour of a naked singularity must be manifest in some observables in the CFT.  Here, we focus only on the simplest of observables -- the one-point functions.

Extracting the CFT signal to a naked singularity is not an easy task.  The numerical evolution of the Gregory-Laflamme instability is computationally costly and has never been done in AdS.  Moreover, extracting the one-point functions requires propagating the likely high-frequency waves from a high-curvature region out to the AdS boundary. Since the naked singularity inevitably develops in finite time, an evolution scheme that uses Cauchy time slices will break down, and the propagation to the boundary must be carried out by other means, such as with outgoing null slices.

As we have seen, the large $D$ limit drastically simplifies the evolution of the instability, but, unfortunately, it complicates signal extraction since the propagation of gravitational waves outside the near-horizon region is strongly suppressed. It is non-perturbative in the large $D$ expansion, and as a result it requires a delicate matching construction (see \cite{Bhattacharyya:2016nhn}).

We therefore turn to a simple model to describe the CFT signal at any finite $D$.  This model correctly predicts the boundary signal from the naked singularity that appears in critical collapse \cite{Chesler:2019ozd}, (i.e the singularity formed by fine-tuning initial data to the edge of black hole formation).  We will adapt this argument to the pinch that arises in the Gregory-Laflamme instability.

Our model relies upon the expected self-similarity of the singularity.  The evolution of the Gregory-Laflamme instability displays a fractal structure, contains numerical hints of a scaling behaviour \cite{Lehner:2010pn,Figueras:2017zwa}, and closely resembles the behaviour of fluid jet instabilities \cite{Cardoso:2006ks}, which are known to give rise to self-similar singularities \cite{Eggers:1995}.  Likewise, the singularities that occur from critical collapse are also self-similar \cite{Choptuik:1992jv}, and indeed the critical solutions have been constructed directly using a self-similar ansatz \cite{Gundlach:1996eg}.

A self-similar function about $t=0$ and $x=0$ is one that satisfies the scaling relation $f(t,x)=e^{-\lambda \Delta_s}f(e^{\lambda}t,e^{\lambda}x)$, where $\Delta_s$ is the scaling dimension.  Here, we will primarily work with dimensionless functions with $\Delta_s=0$.  The function $f$ is called continuously self-similar if the scaling relation is satisfied for all real numbers $\lambda$, and discretely self-similar if it is satisfied only for $\lambda=k\Delta$, for all integers $k$.  In the latter case, $\Delta$ is often called the echoing period in the context of critical collapse.

Note also that if a function is only scale invariant in one variable $t$, we can write $\tau=\log t$ and then by self-similarity $f(\tau,x)=f(\log t, x)=f(\log (e^\lambda t),x)=f(\lambda+\log t,x)=f(\lambda+\tau,x)$. This means that if $f$ is continuously self-similar in $t$, then it is independent of $\tau$, and if $f$ is discretely self-similar in $t$, then it is a periodic function of $\tau$.

\subsection{Self-similar scalar field collapse}

Specializing first to the model of critical collapse, consider a minimally coupled massless scalar field propagating in vacuum (global) AdS, which we write as 
\begin{equation}\label{adsvac}
 ds^2=\frac{1}{\cos^2x}\left(- dt^2+ dx^2+\sin^2x\, d\Omega_{D-2}^2\right)\,.
\end{equation}

Suppose that we start with a scalar field configuration resembling a shell.  The shell will collapse and either form a black hole, or bounce away from the origin.  Fine-tuning the initial data to be between these two scenarios creates a naked singularity that has discrete self-similarity.  The scalar field near this naked singularity will propagate towards the boundary where it will contribute to a vacuum expectation value (VEV) $\langle\mathcal O_\varphi\rangle$.

Let $t_*$ be the time when the naked singularity makes causal contact with the AdS boundary.  Overall, the behavior of $\langle\mathcal O_\varphi\rangle $ contains oscillations which get progressively more rapid, reaching infinity when $t=t_*$.  These oscillations and their dependence on initial data slightly away from the critical point contain information about the critical exponents and echoing period of critical collapse \cite{Chesler:2019ozd}\footnote{Instead of a scalar field, the calculation in \cite{Chesler:2019ozd} uses a particular graviton mode that homogeneously squashes the round sphere, but the behaviour is expected to be the same.}.   These are aspects of the oscillations that are identical to those of asymptotically flat space.

However, \cite{Chesler:2019ozd} found that the overall amplitude of $\langle\mathcal O_\varphi\rangle$ diverges as $1/(t-t_*)$.  As we will argue, this behaviour is a property of AdS, and does not depend on the other characteristics of critical collapse beyond the existence of discrete self-similarity.  

We will assume that the scalar field remains linear and takes a scale-invariant form near $t=0$, $x=0$ where the naked singularity first appears.  Admittedly, this assumption is not entirely justified;  near any singularity, spacetime and any matter fields are highly non-linear.  But we will make this assumption anyway, see what the conclusions are, and then compare the results with those coming from full non-linear evolution.

The spherically-symmetric solution for a massless linear scalar field in AdS can be written as
\begin{equation}
\varphi=\sum_n a_n n^{-\alpha}e^{\pm i\omega_n t}\cos^{D-1}x P_n^{(\alpha,\beta)}[\cos(2x)]\,,
\end{equation}
where $\omega_n=2n+D-1$, $P_n^{(\alpha,\beta)}$ is a Jacobi polynomial, $\alpha=(D-3)/2$, $\beta=(D-1)/2$, and $a_n$ is any combination of mode amplitudes.

Now, we would like to make $\varphi$ a self-similar function in a small region near the origin.  That is, we take $\varphi$ to be invariant under the scaling $t\to\lambda t$, $x\to\lambda x$ for sufficiently high mode numbers.  Using the Mehler–Heine formula
\begin{equation}
\lim_{n\to\infty}n^{-\alpha}P_n^{(\alpha,\beta)}[\cos(2x)]=(nx)^{-\alpha}J_{\alpha}(2nx)\,, \label{MH}
\end{equation}
where $J_\alpha$ is a Bessel function, we see that for high $n$ modes close to the origin $x=0$, the scalar field takes the form
\begin{equation}
\varphi\sim\sum_n a_n F(nt,nx)\,,
\end{equation}
where $F$ stands for some function of purely scaling variables, $nt$ and $nx$, which in our case is given by \eqref{MH}. We see then that scale invariance requires that $a_n = a_{n/\lambda}$.

Now we propagate this linear field to the boundary and read off its signal.  As long as the singularity and its surroundings are sufficiently localised, it is reasonable to assume that the signal mostly propagates linearly on an AdS background.  We can extract the VEV of the dual operator $\mathcal O_\varphi$ from the leading term in the expansion of $\varphi$ near the boundary $x=\pi/2$, to find
\begin{equation}
\langle\mathcal O_\varphi\rangle=\sum_n a_n n^{-\alpha}e^{\pm i\omega_n t}P_n^{(\alpha,\beta)}[-1]\,.
\end{equation}
Using the symmetry relation $P_n^{(\alpha,\beta)}(-z)=(-1)^nP_n^{(\beta,\alpha)}(z)$ and the Mehler–Heine formula again, we see that for high $n$ modes,
\begin{equation}
\langle\mathcal O_\varphi\rangle\sim A\sum_n n a_n e^{\pm i 2n(t-\pi/2)}=\partial_t\sum_n a_n G[n(t-t_*)]\,,
\end{equation}
for some constant $A$ that is independent of $n$. The last equality here is just the observation that the time dependence takes the form $G[n(t-t_*)]$, and the extra factor of $n$ can be obtained by differentiation in $t$.  In other words, the CFT one-point function is the time derivative of a self-similar function about $(t-t_*)$.  If the singularity is discretely self-similar, then the one-point function is the time derivative of a periodic function of $\log(t-t_*)$, so it necessarily has a divergence like
\begin{equation}\label{divO}
 \langle\mathcal O_\varphi\rangle\sim  \frac1{t-t*}\,.
\end{equation}
This is precisely the divergence that was found in a fully non-linear numerical calculation in AdS \cite{Chesler:2019ozd}.  However, if the singularity is continuously self-similar, then this model predicts that the one-point function is (approximately) constant near $t=t_*$. This feature has yet to be compared to any non-linear calculation.

\subsection{Gravitational self-similar solution}

We now apply the same argument to the Gregory-Laflamme instability.

The instability is now driven by the gravitational field, so we will construct a self-similar solution of linearized gravitational perturbations around empty AdS space. The expectation is that our solution reproduces aspects of the self-similar region that are not essentially affected by the presence of a black string horizon around the pinch.\footnote{Observe that there is also a horizon in critical collapse when it is approached from the black hole-forming side, and still our linear model accurately captures the divergent behaviour.}

An important issue is whether the singularity that forms from the Gregory-Laflamme instability is discrete or continuously self-similar. 
However, the nature of the self-similarity remains unclear given the available evidence. 
The observed fractal distribution of small black hole blobs (`satellites') at discrete intervals of the shrinking black string  would suggest discrete self-similarity. 
On the other hand, the thickness of the string horizon appears to follow a continuous linear scaling law \cite{Lehner:2010pn,Figueras:2017zwa}. 
Either type of self-similar behaviour would lead to a violation of cosmic censorship, and it is possible that either or both are present.  

Leaving the ultimate fate of this question aside, we will extract the signal at the boundary coming from a discretely self-similar gravitational singularity. We will solve the gravitational field equations linearized around the AdS vacuum, which we continue to write as \eqref{adsvac}. We seek linear gravitational perturbations that are self-similar near the origin and which fix the boundary metric
\begin{equation}
    ds^2_\partial = -dt^2+d\Omega^2_{D-2}\equiv -dt^2+\gamma_{ij}d\sigma^id\sigma^j\,,
\end{equation}
where we will use latin indices for coordinates on the sphere. The holographic stress-energy tensor will then be extracted from subleading terms in the Fefferman-Graham expansion about the boundary.

\subsubsection*{A heuristic argument}

Before proceeding to the actual calculation, we will present a heuristic argument for the expected behavior of the different components of the holographic stress-energy tensor $\langle T_{\mu\nu}\rangle$, using the fact that it must be traceless and conserved,
\begin{equation}\label{TTstress}
    \langle T^\mu{}_{\mu}\rangle=0\,,\qquad \nabla^\mu\langle T_{\mu\nu}\rangle=0\,.
\end{equation}

Let us  first consider the energy conservation equation,
\begin{equation}\label{econ0}
     \nabla^\mu\langle T_{\mu t}\rangle=-\partial_t \langle T_{t t}\rangle+\nabla^i \langle T_{i t}\rangle =0\,.
\end{equation}
We now recall that discrete self-similarity of correlation functions in the high-frequency limit implies that a time derivative brings a factor of $1/(t-t_*)$. Then, schematically (ignoring all spatial dependence), we must have
\begin{equation}\label{econ}
    \langle T_{t t}\rangle \sim (t-t_*) \langle T_{ti}\rangle\,.
\end{equation}
Similarly, the momentum conservation equation
\begin{equation}\label{mcon0}
     \nabla^\mu\langle T_{\mu i}\rangle=-\partial_t \langle T_{t i}\rangle+\nabla^j \langle T_{j i}\rangle =0
\end{equation}
implies 
\begin{equation}\label{mcon}
    \langle T_{ti}\rangle \sim (t-t_*) \langle T_{ij}\rangle\,.
\end{equation}

These consequences of self-similarity are generic and do not depend on the holographic realization of the system. Now we add to our heuristics the fact that, in the bulk, gravitational tensor perturbations satisfy the same wave equation as a massless scalar field, and therefore we may expect that they diverge in the same manner. Since the spatial (shear) components of the holographic stress-energy tensor are obtained from tensor perturbations, we can then expect that they obey
\begin{equation}\label{tensc}
    \langle T_{ij}\rangle \sim \frac1{t-t_*}\,.
\end{equation}
Taking \eqref{econ}, \eqref{mcon} and \eqref{tensc} together, and adding the tracelessness condition in \eqref{TTstress}, we finally infer that
\begin{equation}\label{scaleT}
    \langle T_{t t}\rangle \sim t-t_*\,,\qquad \langle T_{ti}\rangle\sim \text{const.}\,,\qquad
    \langle T_{ij}\rangle \sim (t-t_*)\gamma_{ij} +\frac1{t-t_*}\pi_{ij}
\end{equation}
where $\pi_{ij}$ denotes a symmetric traceless tensor. Bear in mind that these relations describe only the overall amplitude (the envelope) of functions that oscillate very rapidly in time.

This argument is no more than heuristic, since it applies only between components in the same sector of stress-energy perturbations (i.e., tensors, vectors, or scalars). That is, \eqref{econ} relates the scalar $T_{t t}$ to the scalar-derived components of the vector $T_{ti}$, and \eqref{mcon} relates the vector $T_{t i}$ to the vector-derived components of the tensor $T_{ij}$. The argument we used for \eqref{tensc} applies only to the pure tensors, and not necessarily to the scalar- and vector-derived ones, as we are implicitly assuming to arrive at \eqref{scaleT}. Still, the more detailed analysis that follows does bear out the conclusion \eqref{scaleT}.

\subsubsection*{Scaling solution and holographic stress-energy tensor}

For this linear problem, we can directly use the formalism for gravitational perturbations developed in \cite{Ishibashi_2003,Kodama_2003}. Linear perturbations are decomposed in terms of scalar, vector, and tensor harmonics on the sphere $S^{D-2}$, each with angular mode number $\ell$\footnote{There are multiple spherical harmonics for a given $\ell$ labeled by other mode numbers, but we will suppress these.}. These three sectors of perturbations decouple from each other in the linearized Einstein equation, and the perturbations for each sector $s\in {S,V,T}$ can be reduced to separate ODEs for the master variables $\Phi^{(s)}_\ell(t,r)$ which take the form
\begin{equation}
    \partial_t^2\Phi^{(s)}_\ell=\left(\partial_x^2-\frac{\alpha^2_{\ell}-1/4}{\cos^2x}-\frac{\beta_{(s)}^2-1/4}{\sin^2 x}\right)\Phi^{(s)}_\ell\,,
\end{equation}
where
\begin{equation}
    \alpha_{\ell}=\ell+\frac{D-3}{2}\,,\qquad \beta_{(s)}=\frac{D+1}{2}-c_{(s)}\,,
\end{equation}
with
\begin{equation}
    c_{(T)}=1\,,\qquad c_{(V)}=2\,,\qquad c_{(S)}=3\,.
\end{equation}
In Appendix~\ref{app:gravss}, we give further details needed for our analysis, while for a complete explanation the reader can consult the original works \cite{Ishibashi_2003,Kodama_2003,Ishibashi:2004wx}.

Our general strategy is (1) determine the scaling dimension of the master variables, (2) solve the master equation in terms of Jacobi polynomials and write down a general scale-invariant solution, and (3) extract the holographic stress-energy tensor components. The calculations are presented in Appendix~\ref{app:gravss}, and here we will summarize the main results.

Since the metric has scaling dimension $-2$, we can use dimensional analysis to find the scaling dimension of the master variables.  We then write
\begin{equation}
\Phi^{(s)}_\ell=r^{(D-2)/2}H^{(s)}_\ell\,,
\end{equation}
where $r=\tan x$ is the usual radial coordinate, and $H^{(s)}_\ell$ has scaling dimension $0$.  Then, solving the master equation and summing over modes, we have
\begin{equation}
    H^{(s)}(t,x)=\sum_{n,\ell}a_{n\ell}\,n^{\ell-\alpha_{\ell}}e^{-i\omega^{(s)}_{n\ell} t}\sin^\ell x\cos^{D-c_{(s)}}x P^{(\alpha_{\ell},\beta_{(s)})}_n[\cos(2x)]\,,
\end{equation}
where
\begin{equation}
    \omega^{(s)}_{n\ell}=2n+\ell+D-c_{(s)}\,.
\end{equation}

As in the scalar field case, we can use the Mehler–Heine formula \eqref{MH} to see that for high $n$ modes (assuming also $n\gg \ell$) near the origin $x=0$\,,
\begin{equation}
    H^{(s)}(t,x)\sim\sum_{n,\ell}a_{n\ell}\,e^{-i2nt}(nx)^{\ell}(nx)^{-\alpha_{\ell}}J_{\alpha_{\ell}}(2nx)=\sum_{n,\ell}a_{n\ell} F(nt,nx)\,,
\end{equation}
for some function $F$ of scaling variables $nt$ and $nx$.  We again choose $a^{(s)}_n = a^{(s)}_{n/\lambda}$ to preserve scale invariance.

At the boundary $x=\pi/2$ for high $n$ modes, we have
\begin{equation}
    H^{(s)}(t,x)\sim\sum_{n,\ell}A_\ell a_{n\ell}\,\left[n^{2-c_{(s)}}\right]\,e^{-i\omega^{(s)}_{n\ell} t}\cos^{D-c_{(s)}}x\left[1+\ord{\frac{\pi}{2}-x}\right]\,,
\end{equation}
for some constants $A_\ell$ independent of $n$.  The factor of $n^{2-c_{(s)}}$ that we highlighted in brackets is largely responsible for the different behaviours that we will see in the components of the boundary stress-energy tensor.

By relating the master variables to the metric components and continuing the series to higher order in a Fefferman-Graham expansion, we can eventually extract the boundary stress-energy tensor.  The details of this calculation are given in the appendix, where we also verify that the stress-energy tensor satisfies \eqref{TTstress}. 

For tensor perturbations, the boundary stress-energy tensor takes the form
\begin{align}
    \langle T_{ij}\rangle&\sim \sum_{n,\ell} a_{n\ell}\, n\, G_\ell[n(t-t_*)]\mathbb T^{(\ell)}_{ij}\\
    &\sim \frac1{t-t_*}\mathbb{T}_{ij}\,,
\end{align}
where $\mathbb{T}_{ij}$ are tensor harmonics of $S^{D-2}$. These are divergent fluctuations in the shear stresses of the CFT.  In this case, the factor $n^{2-c_{(T)}}=n$ directly contributes to the divergence.

For vector perturbations, we have
\begin{align}
    \langle T_{ti}\rangle&\sim \sum_{n,\ell} a_{n\ell}G^{(1)}_\ell[n(t-t_*)]\mathbb V_i\sim \mathbb V_i\,,\\
    \langle T_{ij}\rangle&\sim\sum_{n,\ell}a_{n\ell}\, n\, G^{(2)}_\ell[n(t-t_*)]D_{(i}\mathbb{V}_{j)}\sim \frac{1}{t-t_*}D_{(i}\mathbb{V}_{j)}\,,
\end{align}
where $\mathbb{V}_{i}$ are vector harmonics of $S^{D-2}$.  

The factor $n^{2-c_{(V)}}=1$ in the master variables causes the momentum components to have constant amplitude.  Once we have this behaviour for the momentum components of vector perturbations, the remaining components are fixed by momentum conservation \eqref{mcon0}.  As a result, the shear stresses diverge for these perturbations.

Finally, for scalar perturbations, we have
\begin{align}
    \langle T_{tt}\rangle&\sim \sum_{n,\ell} \frac{1}{n}a_{n\ell}G^{(0)}_\ell[n(t-t_*)]\mathbb S\sim (t-t_*)\mathbb S\,,\\
    \langle T_{ti}\rangle&\sim \sum_{n,\ell} a_{n\ell}G^{(1)}_\ell[n(t-t_*)]\mathbb S_i\sim \mathbb S_i\,,\\
    \langle T_{ij}\rangle&\sim\sum_{n,\ell}a_{n\ell}\left[\frac{1}{n} G^{(L)}_\ell[n(t-t_*)]\gamma_{ij}\mathbb S+n\,G^{(T)}_\ell[n(t-t_*)]\mathbb S_{ij}\right]\\
    &\sim (t-t_*)A^{(L)}\gamma_{ij}\mathbb S+\frac{1}{t-t_*}A^{(T)}\mathbb S_{ij}\,,
\end{align}
for some amplitudes $A^{(L)}$, $A^{(T)}$, and where $\mathbb{S}$ are scalar harmonics of $S^{D-2}$, and $\mathbb S_i$, $\mathbb S_{ij}$ are the associated vector- and tensor-derived scalar harmonics.  

The factor $n^{2-c_{(S)}}=1/n$ in the master variables causes the energy component to vanish linearly.  Once the energy density in the stress-energy tensor is given for scalar perturbations, the behaviour of the remaining components is determined by the tracelessness and conservation of the stress-energy tensor \eqref{TTstress}.  We see that just like the vectors and tensors, the shear stresses diverge and the momentum components have constant amplitude.  The pressures, like the energy density, vanish as $t-t_*$.

So we see that all sectors contribute to a $1/(t-t_*)$ divergence in the trace-free part of the spatial components, associated with shear stresses.  But the amplitude of the signal remains constant for the momentum components, and vanishes as $(t-t_*)$ for the energy density and pressures.  Nevertheless, in all components the boundary signal is not smooth: it oscillates an infinite number of times before the singularity occurs and therefore reaches arbitrarily high frequencies. In the quantum CFT at finite $N$, we expect to observe a few quanta with energies that scale with $N^2$. Then, the stress-energy conservation laws imply large shears $\sim N^2$, in a pattern on the boundary sphere whose angular distribution is presumably dominated by low multipoles $\ell\gtrsim 2$. 

\section{Outlook}
\label{sec:outlook}

The AdS/CFT correspondence still has to throw clearer light on the dual (microscopic and fundamental) meaning of known violations of cosmic censorship: What mechanisms drive them? How do they manifest in the  dual large-$N$ field theory? How does the theory deal with them when $N$ is finite and quantum gravitational bulk effects are incorporated? In this article, we have aimed at making progress in the first two questions, and hope that this will contribute to eventually answering the third one.

The bulk dynamics that we have found is very rich. We have identified several qualitatively new aspects of the evolution of unstable black strings, and of the way naked singularities appear. Perhaps the most peculiar one is the `delayed censorship' where a singular pinch forms and is subsequently washed out by a black tsunami. Another salient aspect is the characteristic signal in the holographic stress-energy tensor that is dual to our model of self-similar singularity formation.

The large-$D$ methods have been instrumental to our analysis, as they are very efficient for showing how a process is triggered that non-linearly drives a horizon to pinch. However, they break down before entering the crucial self-similar regime of evolution. For this, other methods, including dedicated numerical simulations, will be required. Nevertheless, since the leading order large-$D$ theories have a built-in horizon regulator, they may also manage to capture the evolution \emph{after} the singularity is resolved. This prospect is especially appealing if the exit from the singularity is controlled by a post-pinch attractor, as is the case in fluid-jet pinch-offs. For the breakup of black strings, the existence of these attractors remains conjectural.

We have argued that the formation of the singularity leaves a distinctive mark on the one-point functions of field theory operators. However, this argument required that the singularity be discrete self-similar.  When the approach to the singularity is continuously self-similar, these observables will remain constant and not give any detectable signal of the event in the bulk. It would be very interesting to investigate the system with other probes, such as higher-point functions or non-local observables. For instance, bulk geodesics that begin and end at the boundary approximate two-point functions, and they seem likely to give a signal when they pass close to the self-similar region. These effects could be investigated in our linearized scaling solution.

The model predicts that the high-frequency signal issuing from a discrete self-similar singularity should have holographic energy density vanishing as $t-t_*$, and shear components that diverge as $1/(t-t_*)$, which perhaps would have been difficult to anticipate from the CFT side.  A main challenge is to formulate a gauge-theoretical resolution of the divergence at finite $N$ (or possibly finite 't~Hooft coupling $\lambda$, if string theory describes the bulk).  It seems plausible that singularity formation can be mapped to a dual problem in matrix quantum mechanics, perhaps within the BFSS/D0 model, or some other suitable matrix theory\footnote{For a related but slightly different approach, see \cite{Hanada:2018qpf}.}. If that is possible, one might be able to see how the finite $N$ theory resolves the singularity -- that is, how the system evolves past the moment when certain matrix operators acquire large values that scale with $N$. Our results give a prediction for how the operator VEVs will behave in the approach to this enigmatic moment.

\section*{Acknowledgments}

We are grateful to Tom\'as Andrade, Gary Horowitz, and Toby Wiseman for interesting discussions and comments on a draft of the article. Work supported by ERC Advanced Grant GravBHs-692951, MICINN grant PID2019-105614GB-C22, AGAUR grant 2017-SGR 754, and State Research Agency of MICINN through the ``Unit of Excellence María de Maeztu 2020-2023” award to the Institute of Cosmos Sciences (CEX2019-000918-M). RS is supported by JSPS KAKENHI Grant Number JP18K13541 and partly by Osaka City University Advanced Mathematical Institute (MEXT Joint Usage/Research Center on Mathematics and Theoretical Physics). MT is also supported by the European Research Council (ERC) under the European Union’s Horizon 2020 research and innovation programme (grant agreement No 852386). DL is supported by a Minerva Fellowship of the Minerva Stiftung Gesellschaft fuer die Forschung mbH.

\appendix
\section{Analysis of the boundary signal}\label{app:gravss}
Here, we put the technical details of our calculation that extracts the boundary signal to a self-similar gravitational singularity.  This calculation involves considering linear gravitational perturbations on the background \eqref{adsvac}, which we reproduce here as
\begin{equation}
 ds^2=\frac{1}{\cos^2x}\left(- dt^2+ dx^2+\sin^2x\, d\Omega_{D-2}^2\right)\equiv g_{ab}(y)dy^ady^b+r^2(y)\gamma_{ij}d\sigma^id\sigma^j\,,
\end{equation}
where we will use middle latin letters ($i$, $j$, etc.) for coordinates on the sphere, and lower-alphabetical indices ($a$, $b$, etc.) for the remaining two coordinates.  Here, 
\begin{equation}
    r=\tan x
\end{equation}
is the radial coordinate.  For this section, we will use $\nabla$ for the covariant derivative on $g_{ab}$, and $\hat\nabla$ for the covariant derivative on $\gamma_{ij}$.

\subsection{Master Equation for Gravitational Perturbations}
Following the formalism from \cite{Ishibashi_2003,Kodama_2003,Ishibashi:2004wx}, gravitational perturbations can be separated into tensor, vector, and scalar harmonics on the sphere.  The perturbations then reduce to a set of decoupled master equations.  The tensor perturbations take the form
\begin{equation}
    \delta g_{ab}=0\,,\qquad \delta g_{ai}=\delta g_{ia}=0\,,\qquad \delta g_{ij}=2r^2H_T(y)\mathbb T_{ij}\,,
\end{equation}
where $\mathbb T_{ij}$ are tensor harmonics on the sphere that satisfy
\begin{equation}
    (\hat\nabla^2+k_T^2)\mathbb T_{ij}=0\,,\qquad \mathbb T^i{}_{i}=0\,,\qquad \hat\nabla^i\mathbb T_{ij}=0\,,
\end{equation}
\begin{equation}
    k_T^2=\ell(\ell+D-3)-2\,.
\end{equation}
The function $H_T$ can be related to a master variable as
\begin{equation}
    H_T=r^{-\frac{D-2}{2}}\Phi^{(T)}\,.
\end{equation}

The vector perturbations take the form
\begin{equation}
    \delta g_{ab}=0\,,\qquad \delta g_{ai}=\delta g_{ia}=r f_a(y)\mathbb V_i\,,\qquad \delta g_{ij}=-\frac{2}{k_V}r^2H^{(1)}_T(y)\hat\nabla_{(i}\mathbb V_{j)}\,,
\end{equation}
where $\mathbb V_{i}$ are vector harmonics on the sphere that satisfy
\begin{equation}
    (\hat\nabla^2+k_V^2)\mathbb V_{i}=0\,,\qquad \hat\nabla^i\mathbb V_i=0\,,
\end{equation}
\begin{equation}
    k_V^2=\ell(\ell+D-3)-1\,.
\end{equation}
The gauge-invariant combinations of $f_a$ and $H^{(1)}_T$ can be related to a master variable via
\begin{equation}
    F_a\equiv f_a+\frac{r}{k_V}\nabla_aH^{(1)}_T=\frac{1}{r^{D-3}}\epsilon_{ab}\nabla^b(r^{-\frac{D-2}{2}}\Phi^{(V)})\,.
\end{equation}
Given $\Phi^{(V)}$, $H^{(1)}_T$ can be freely chosen to fix a gauge for determining the metric components.

Finally, the scalar perturbations take the form
\begin{equation}
    \delta g_{ab}=f_{ab}\mathbb S\,,\qquad \delta g_{ai}=\delta g_{ia}=r f^{(1)}_a(y)\mathbb S_i\,,\qquad \delta g_{ij}=2r^2\left[H_L(y)\gamma_{ij}\mathbb S+H^{(2)}_T(y)\mathbb S_{ij}\right]\,,
\end{equation}
where $\mathbb S$ are scalar harmonics on the sphere that satisfy
\begin{equation}
    (\hat\nabla^2+k_S^2)\mathbb S=0\,,
\end{equation}
\begin{equation}
    k_S^2=\ell(\ell+D-3)\,,
\end{equation}
and the vector and tensor-derived scalar harmonics are defined by
\begin{equation}
    \mathbb S_i=-\frac{1}{k_S}\hat\nabla_i\mathbb S\,,\qquad \mathbb S_{ij}=\frac{1}{k_S^2}\hat\nabla_i\hat\nabla_j\mathbb S+\frac{1}{D-2}\gamma_{ij}\mathbb S\,.
\end{equation}
We can now define the following gauge-invariant combinations of $f_{ab}$, $f^{(1)}_a$, $H_L(y)$, and $H^{(2)}_T$
\begin{align}
    F_{ab}&=f_{ab}+\nabla_aX_b+\nabla_bX_a\,,\\
    F&=H_L+\frac{1}{D-2}H^{(2)}_T+\frac{1}{r}(\nabla^ar) X_a\,,
\end{align}
where
\begin{equation}
    X_a=\frac{r}{k_S}\left(f^{(1)}_a+\frac{r}{k_S}\nabla_aH^{(2)}_T\right)\,.
\end{equation}
These gauge-invariant combinations can be obtained from a master variable via
\begin{align}
    F_{ab}+(D-4)F g_{ab}&=\frac{1}{r^{D-4}}\left[\nabla_a\nabla_b-\frac{1}{2}g_{ab}\nabla^2\right]\Big(r^{-\frac{D-2}{2}}\Phi^{(S)}\Big)\,,\\
    F&=\frac{1}{4r^2}\left[2r(\nabla_ar)\nabla^a+2\left(\frac{k_S^2}{D-2}-1\right)-2r^2\right]\Big(r^{-\frac{D-2}{2}}\Phi^{(S)}\Big)\,.
\end{align}
Given $\Phi^{(S)}$ and a choice of $X_a$ and $H^{(2)}_T$, the full metric components can be obtained.

The linearised Einstein equation gives a differential equation for the master variables, which can be written as \cite{Ishibashi:2004wx}
\begin{equation}
    \partial_t^2\Phi^{(s)}_\ell=\left(\partial_x^2-\frac{\alpha^2_{\ell}-1/4}{\cos^2x}-\frac{\beta_{(s)}^2-1/4}{\sin^2 x}\right)\Phi^{(s)}_\ell\,,
\end{equation}
where
\begin{equation}
    \alpha_{\ell}=\ell+\frac{D-3}{2}\,,\qquad \beta_{(s)}=\frac{D+1}{2}-c_{(s)}
\end{equation}
with
\begin{equation}
    c_{(T)}=1\,,\qquad c_{(V)}=2\,,\qquad c_{(S)}=3\,.
\end{equation}

\subsection{Scaling Solution}
Now we seek solutions to this equation that preserve the boundary metric
\begin{equation}
    ds^2_\partial=-dt^2+d\Omega_{D-2}^2\,,
\end{equation}
and are scale-invariant near the origin for high radial mode numbers $n$.

We note that the metric has scaling dimension $-2$.  Then using dimensional analysis, we find that the master variables can be written as
\begin{equation}
\Phi^{(s)}_\ell=r^{(D-2)/2}H^{(s)}_\ell,
\end{equation}
with $H^{(s)}_\ell$ having scaling dimension $0$.  Solving the master equation and summing over modes, we have
\begin{equation}\label{linsol}
    H^{(s)}(t,x)=\sum_{n,\ell}a_{n\ell}\,n^{\ell-\alpha_{\ell}}e^{-i\omega^{(s)}_{n\ell} t}\sin^\ell x\cos^{D-c_{(s)}}x P^{(\alpha_{\ell},\beta_{(s)})}_n[\cos(2x)]\,,
\end{equation}
\begin{equation}
    \omega^{(s)}_{n\ell}=2n+\ell+D-c_{(s)}\,.
\end{equation}
Using the Mehler–Heine formula \eqref{MH} we see that high $n$ modes (with $n\gg \ell$) near the origin $x=0$ satisfy
\begin{equation}
    H^{(s)}(t,x)\sim\sum_{n,\ell}a^{(s)}_{n\ell}e^{-i2nt}(nx)^{\ell}(nx)^{-\alpha_{\ell}}J_{\alpha_{\ell}}(2nx)=\sum_{n,\ell}a_{n\ell} F(nt,nx)\,,
\end{equation}
for some function $F$ of scaling variables $nt$ and $nx$.  We can choose $a^{(s)}_n = a^{(s)}_{n/\lambda}$ to preserve scale invariance.

\subsection{Stress-energy tensor}
Now we wish to extract the boundary stress-energy tensor from this solution.  In doing so, we also generalise the calculation in \cite{Friess:2006kw} to higher dimensions.  The stress-energy tensor can be obtained via a Fefferman-Graham expansion about the AdS boundary \cite{deHaro:2000vlm}, of the form
\begin{equation}
ds^2=\frac{1}{z^2}\left[dz^2+ds^2_\partial+\sum_kds^2_{(k)}z^k+z^{\frac{D-1}{2}}\log z\sum_k\widetilde ds^2_{(k)} z^k\right]\,,
\end{equation}
where the boundary metric $ds^2_\partial$ and each term $ds^2_{(k)}$ and $\widetilde ds^2_{(k)}$ are independent of the radial coordinate $z$.  The log terms are only present when $D$ is odd.

For our metric perturbations, we will find that
\begin{equation}
ds^2=ds^2_{AdS}+\frac{1}{z^2}\left[\delta g_{\mu\nu}dx^\mu dx^\nu z^{(D-1)}+\ldots\right]\,,
\end{equation}
where we are using Greek indices for boundary coordinates.  In this case, the boundary stress-energy tensor relative to that of AdS can be written
\begin{equation}
    \langle T_{\mu\nu}\rangle=\langle T_{\mu\nu}\rangle_{\mathrm{Full}}-\langle T_{\mu\nu}\rangle_{AdS}=\frac{D-1}{16\pi G_D}\delta g_{\mu\nu}\,.
\end{equation}

Our task is to take our solution from \eqref{linsol}, perform a boundary expansion, obtain the metric in Fefferman-Graham gauge, and read off the stress-energy tensor.  The stress-energy tensor is always traceless and conserved,
and we will check this explicitly in our calculation.

\subsubsection{Tensor Perturbations}
For the tensor perturbations, the solution is already in Fefferman-Graham gauge, so we can read off the stress-energy tensor immediately by just evaluating the solution at the boundary.  We can use
\begin{equation}
P^{(\alpha,\beta)}_n[-1]=(-1)^n\frac{\Gamma(n+1+\beta)}{\Gamma(n+1)\Gamma(1+\beta)}
\end{equation}
to get
\begin{equation}
\langle T_{ij}\rangle=\frac{D-1}{8\pi G_D}\sum_{n,\ell}a_{n\ell}(-1)^nn^{\ell-\alpha_{\ell}}\frac{\Gamma(n+1+\beta_{(T)})}{\Gamma(n+1)\Gamma(1+\beta_{(T)})}e^{-i\omega^{(T)}_{n\ell} t}\mathbb T_{ij}
\end{equation}
The stress-energy tensor is already traceless and conserved by the properties of $\mathbb T_{ij}$.

When $n$ is large, we have $\omega^{(T)}_{n\ell}\sim 2n$ and we can use
\begin{equation}
\lim_{n\to\infty}\frac{\Gamma(n+a)}{\Gamma(n+b)}=n^{a-b}
\end{equation}
to get
\begin{equation}
\langle T_{ij}\rangle\sim\frac{D-1}{8\pi G_D}\sum_{n,\ell}a_{n\ell}\,n \,e^{i 2n(t-\pi/2)}\mathbb T_{ij}\sim\frac{1}{t-t_*}\mathbb T_{ij}\,.
\end{equation}

\subsubsection{Vector Perturbations}
For the vector perturbations, we must choose $H^{(1)}_T$ so that $f_r=0$.  As an expansion, we find
\begin{equation}
    \Phi^{(V)}_\ell(t,r)=\sum_{n,\ell}A^{(V)}_{n\ell}(t)\frac{1}{r^{\frac{D-2}{2}}}\left(1+\frac{k_V^2-(D-3)-(\omega^{(V)}_{n\ell})^2}{2(D-1)}\frac{1}{r^2}+\ldots\right)\,,
\end{equation}
where
\begin{equation}\label{Afactor}
    A^{(s)}_{n\ell}(t)=a_{n\ell}\,(-1)^nn^{\ell-\alpha_{\ell}}\frac{\Gamma(n+1+\beta_{(s)})}{\Gamma(n+1)\Gamma(1+\beta_{(s)})}e^{-i\omega^{(s)}_{n\ell} t}\,.
\end{equation}
To move to Fefferman-Graham gauge in an expansion in $1/r$, we choose
\begin{equation}
    H^{(1)}_T(t,r)=\sum_{n,\ell}i\frac{\omega^{(V)}_{n\ell} k_V}{D-1}A^{(V)}_{n\ell}(t)\frac{1}{r^{D-1}}+\ldots
\end{equation}
to find
\begin{align}
    \langle T_{ti}\rangle&=\frac{1}{16\pi G_D}\sum_{n,\ell}A^{(V)}_{n\ell}(t)\left[D-3-k_V^2\right]\mathbb V_i\,,\\
    \langle T_{ij}\rangle&=\frac{1}{16\pi G_D}\sum_{n,\ell}A^{(V)}_{n\ell}(t)\left[-2i\omega^{(V)}_{n\ell}\right]\hat\nabla_{(i}\mathbb V_{j)}\,.
\end{align}
Tracelessness and $\nabla^\mu\langle T_{\mu t}\rangle=0$ follow immediately from $\nabla^i\mathbb V_i=0$.  For the remaining components, we have
\begin{equation}
    \nabla^\mu\langle T_{\mu i}\rangle=-\partial_t\langle T_{t i}\rangle+\hat\nabla^j\langle T_{ji}\rangle=\frac{1}{16\pi G_D}\sum_{n,\ell}i\omega^{(V)}_{n\ell}A^{(V)}_{n\ell}(t)\left[(D-3-k_V^2)\mathbb V_i-2\hat\nabla^j\hat\nabla_{(j}\mathbb V_{i)}\right]\,.
\end{equation}
But using the fact that the Ricci tensor for the sphere satisfies $\hat R_{ij}=(D-3)\gamma_{ij}$, we have
\begin{align}
    2\hat\nabla^j\hat\nabla_{(j}\mathbb V_{i)}&=\hat\nabla^2\mathbb V_i+\hat\nabla^j\hat\nabla_i\mathbb V_j\\
    &=-k_V^2\mathbb V_i+[\hat\nabla^j\hat\nabla_i-\hat\nabla_i\hat\nabla^j]\mathbb V_j\\
    &=-k_V^2\mathbb V_i+\hat R_i{}^j\mathbb V_j\\
    &=(D-3-k_V^2)\mathbb V_i\,,
\end{align}
which gives us $\nabla^\mu\langle T_{\mu i}\rangle=0$.

Now at large mode number, we have $\omega^{(V)}_{n\ell}\sim 2n$ and
\begin{equation}
    A^{(V)}_{n\ell}(t)\sim a_{n\ell}\,e^{i 2n(t-\pi/2)}\,,
\end{equation}
which implies
\begin{align}
    \langle T_{ti}\rangle&\sim\frac{1}{16\pi G_D}\sum_{n,\ell}a_{n\ell}\,e^{i 2n(t-\pi/2)}\left[D-3-k_V^2\right]\mathbb V_i\sim \mathbb V_i\,,\\
    \langle T_{ij}\rangle&\sim\frac{1}{16\pi G_D}\sum_{n,\ell}-4ia_{n\ell}\,n\,e^{i 2n(t-\pi/2)}\hat\nabla_{(i}\mathbb V_{j)}\sim \frac{1}{t-t_*}\hat\nabla_{(i}\mathbb V_{j)}\,.
\end{align}

\subsubsection{Scalar Perturbations}
As before, we expand the master variables
\begin{align}
    \Phi^{(S)}_\ell(t,r)=\sum_{n,\ell}&A^{(S)}_{n\ell}(t)\frac{1}{r^{\frac{D-4}{2}}}\bigg(1+\frac{k_S^2-(\omega^{(S)}_{n\ell})^2}{2(D-3)}\frac{1}{r^2}+\\
    &\qquad+\frac{[k_S^2-(\omega^{(S)}_{n\ell})^2]^2-2(D-1)k_S^2+4(D-2)(\omega^{(S)}_{n\ell})^2}{8(D-3)(D-1)}+\ldots\bigg)\,,
\end{align}
where $A^{(S)}_{n\ell}(t)$ is given by \eqref{Afactor}.

To move to Fefferman-Graham gauge in a $1/r$ expansion, we choose
\begin{align}
    X_t(t,r)&=\sum_{n,\ell}-i \omega^{(S)}_{n\ell}A^{(S)}_{n\ell}(t)\frac{1}{r^{D-3}}\left(\frac{k_S^2-(D-2)(\omega^{(S)}_{n\ell})^2+2(D-3)[k_S^2-(D-2)]}{2(D-3)(D-2)(D-1)}+\ldots\right)\\
    X_r(t,r)&=-\sum_{n,\ell}A^{(S)}_{n\ell}(t)\frac{1}{r^{D-2}}\frac{k_S^2-(D-2)(\omega^{(S)}_{n\ell})^2}{2(D-3)(D-2)}\left(1+\frac{k_S^2-4(D-2)-(\omega^{(S)}_{n\ell})^2}{2(D-1)}\frac{1}{r^2}+\ldots\right)\\
    H^{(2)}_T(t,r)&=\sum_{n,\ell}A^{(S)}_{n\ell}(t)\frac{1}{r^{D-1}}\left(\frac{k_S^2\Big[k_S^2-(D-2)(\omega^{(S)}_{n\ell})^2\Big]}{2(D-3)(D-2)(D-1)}+\ldots\right)
\end{align}
to find
\begin{align}
    \langle T_{tt}\rangle&=\frac{1}{16\pi G_D}\sum_{n,\ell}A^{(S)}_{n\ell}(t)\left[\frac{k_S^2[k_S^2-(D-2)]}{D-2}\right]\mathbb S\,,\\
    \langle T_{ti}\rangle&=\frac{1}{16\pi G_D}\sum_{n,\ell}A^{(S)}_{n\ell}(t)\left[-i\,\omega^{(S)}_{n\ell}\frac{k_S[k_S^2-(D-2)]}{D-2}\right]\mathbb S_i\,,\\
    \langle T_{ij}\rangle&=\frac{1}{16\pi G_D}\sum_{n,\ell}A^{(S)}_{n\ell}(t)\left\{\left[\frac{k_S^2[k_S^2-(D-2)]}{(D-2)^2}\right]\gamma_{ij}\mathbb S+\left[\frac{k_S^2[k_S^2-(D-2)(\omega^{(S)}_{n\ell})^2]}{(D-3)(D-2)}\right]\mathbb S_{ij}\right\}\,.
\end{align}
The tracelessness of the stress-energy tensor follow from $\mathbb S^i{}_i=0$ and $\delta^{i}{}_i=D-2$.  To check that the stress-energy tensor is conserved, we first compute
\begin{equation}
\nabla^\mu\langle T_{\mu t}\rangle=\frac{1}{16\pi G_D}\sum_{n,\ell}i \omega^{(S)}_{n\ell}A^{(S)}_{n\ell}(t)\left[\frac{k_S^2[k_S^2-(D-2)]}{D-2}\mathbb S-\frac{k_S[k_S^2-(D-2)]}{D-2}\hat\nabla^i\mathbb S_i\right]\,,
\end{equation}
which vanishes because of the identity $\hat\nabla^i\mathbb S_i=k_S\mathbb S$\,.  And for the remaining component,
\begin{align}
\nabla^\mu\langle T_{\mu i}\rangle=\frac{1}{16\pi G_D}\sum_{n,\ell}A^{(S)}_{n\ell}(t)&\bigg[(\omega^{(S)}_{n\ell})^2\frac{k_S[k_S^2-(D-2)]}{D-2}\mathbb S_i+\frac{k_S^2[k_S^2-(D-2)]}{(D-2)^2}\hat\nabla_i\mathbb S+\\
&\qquad+\frac{k_S^2[k_S^2-(D-2)(\omega^{(S)}_{n\ell})^2]}{(D-3)(D-2)}\hat\nabla^j\mathbb S_{ji}\bigg]\,,
\end{align}
which vanishes by using the identities $\hat\nabla_i\mathbb S=-k_S\mathbb S_i$ and $\hat\nabla^j \mathbb S_{ji}=\frac{D-3}{D-2}\frac{k_S^2-(D-2)}{k_S}\mathbb S_i$.

At large $n$, we again have $\omega^{(S)}_{n\ell}\sim 2n$ and
\begin{equation}
    A^{(S)}_{n\ell}(t)\sim a_{n\ell}\frac{1}{n}e^{i 2n(t-\pi/2)}\,,
\end{equation}
which implies
\begin{align}
    \langle T_{tt}\rangle&\sim\frac{1}{16\pi G_D}\sum_{n,\ell}a_{n\ell}\,e^{i 2n(t-\pi/2)}\frac{1}{n}\left[\frac{k_S^2[k_S^2-(D-2)]}{D-2}\right]\mathbb S\sim (t-t_*)\mathbb S\,,\\
    \langle T_{ti}\rangle&\sim\frac{1}{16\pi G_D}\sum_{n,\ell}a_{n\ell}\,e^{i 2n(t-\pi/2)}\left[-2i\frac{k_S[k_S^2-(D-2)]}{D-2}\right]\mathbb S_i\sim \mathbb S_i\,,\\
    \langle T_{ij}\rangle&\sim\frac{1}{16\pi G_D}\sum_{n,\ell}-4ia_{n\ell}\,n\,e^{i 2n(t-\pi/2)}\left\{\frac{1}{n}\left[\frac{k_S^2[k_S^2-(D-2)]}{(D-2)^2}\right]\gamma_{ij}\mathbb S+n\left[-\frac{4k_S^2}{D-3}\right]\mathbb S_{ij}\right\}\\
    &\sim (t-t_*)A^{(L)}\gamma_{ij}\mathbb S+\frac{1}{t-t_*}A^{(T)}\mathbb S_{ij}\,,
\end{align}
for some amplitudes $A^{(L)}$, $A^{(T)}$.

\end{document}